\documentclass[review]{elsarticle}

\usepackage{amsmath}
\usepackage{amsfonts}
\usepackage{amssymb}
\usepackage{amsthm}
\usepackage{bm}
\usepackage{indentfirst}
\usepackage{graphicx}
\usepackage{subfigure}
\usepackage{array}
\usepackage{fullpage}

\DeclareMathAlphabet{\mathsfsl}{OT1}{cmss}{m}{sl}

\newcommand{\PreserveBackslash}[1]{\let\temp=\\#1\let\\=\temp}
\newcolumntype{C}[1]{>{\PreserveBackslash\centering}p{#1}}
\newcolumntype{R}[1]{>{\PreserveBackslash\raggedleft}p{#1}}
\newcolumntype{L}[1]{>{\PreserveBackslash\raggedright}p{#1}}

\numberwithin{equation}{section}

\theoremstyle{definition}

\usepackage{anyfontsize}
\usepackage{enumerate}
\usepackage{esint}
\usepackage{etoolbox}
\usepackage{isomath}
\usepackage{mathrsfs}
\usepackage{mathtools}
\usepackage{multicol}
\usepackage{pgfplots}
\usepackage{scalerel}
\usepackage{stackengine}
\usepackage{tabulary}
\usepackage{tikz}

\makeatletter
\newcommand*\bdot{\mathpalette\bdot@{.65}}
\newcommand*\bdot@[2]{\mathbin{\vcenter{\hbox{\scalebox{#2}{$\m@th#1\bullet$}}}}}
\makeatother

\makeatletter
\newcommand*\bddot{\mathpalette\bddot@{.65}}
\newcommand*\bddot@[2]{\mathbin{\vcenter{\hbox{\scalebox{#2}
    {$\m@th#1\smash{{}_{\bullet}^{\bullet}}$}}}}}
\makeatother

\newcommand{\circled}[2][]{%
  \tikz[baseline=(char.base)]{%
    \node[shape = circle, draw, inner sep = .5pt]
    (char) {\phantom{\ifblank{#1}{#2}{#1}}};%
    \node at (char.center) {\makebox[0pt][c]{#2}};}}
\robustify{\circled}

\def\dist{\operatorname{dist}}

\makeatletter
\newcommand{\opnorm}{\@ifstar\@opnorms\@opnorm}
\newcommand{\@opnorms}[1]{%
  \left|\mkern-1.5mu\left|\mkern-1.5mu\left|
   #1
  \right|\mkern-1.5mu\right|\mkern-1.5mu\right|
}
\newcommand{\@opnorm}[2][]{%
  \mathopen{#1|\mkern-1.5mu#1|\mkern-1.5mu#1|}
  #2
  \mathclose{#1|\mkern-1.5mu#1|\mkern-1.5mu#1|}
}
\makeatother

\stackMath
\newcommand\reallywidecheck[1]{%
\savestack{\tmpbox}{\stretchto{%
  \scaleto{%
    \scalerel*[\widthof{\ensuremath{#1}}]{\kern-.6pt\bigwedge\kern-.6pt}%
    {\rule[-\textheight/2]{1ex}{\textheight}}
  }{\textheight}%
}{0.5ex}}%
\stackon[1pt]{#1}{\scalebox{-1}{\tmpbox}}%
}

\usepackage[utf8]{inputenc}
\usepackage{bm}
\usepackage{mathrsfs}
\usepackage{amsmath}
\usepackage{amsfonts}
\usepackage{amsthm}
\usepackage{color}
\usepackage{setspace}
\usepackage{float}
\usepackage{import}
\usepackage{url}
\usepackage{multicol}
\usepackage{subfigure}
\biboptions{sort&compress}
\usepackage[top=1in,bottom=1in,left=1in,right=1in]{geometry}
\newcommand{\real}{\mathbb{R}}

\newcommand{\R}{\mathbb{R}}
\newcommand{\C}{\mathbb{C}}

\newcommand{\mcL}{\mathcal{L}}

\newcommand{\mcA}{\mathcal{A}}

\newcommand{\mcP}{\mathcal{P}}

\usepackage{xcolor}

\def\omg{{\Omega}}
\def\omgi{\mathcal{I}{\Omega}}
\def\omgb{\mathcal{B}\Omega}
\def\omgbb{\mathcal{B}\mathcal{B}\Omega}

\def \fb{\bm{f}}

\def \gb{\bm{g}}
\def \ub{\bm{u}}

\def \wb{\bm{w}}
\def \vb{\bm{v}}
\def \xb{\bm{x}}
\def \Rb{\bm{R}}
\def \Vb{\bm{V}}

\def \zb{\bm{z}}

\def \nb{\bm{n}}
\def \yb{\bm{y}}

\def \eb{\bm{e}}

\def \xib{{\boldsymbol\xi}}

\newcommand{\vertii}[1]{{\left\vert\left\vert #1
    \right\vert\right\vert}}    
    
\newcommand{\verti}[1]{{\left\vert #1
    \right\vert}}  
    
\newtheorem{theorem}{Theorem}
\newtheorem{corollary}{Corollary}
\newtheorem{lemma}{Lemma}
\newtheorem{remark}{Remark}
   
\newtheorem{assumption}{Assumption}

\begin{document}

\begin{frontmatter}

\title{A Meshfree Peridynamic Model for Brittle Fracture\\ in Randomly Heterogeneous Materials}

\address[yy]{Department of Mathematics, Lehigh University, Bethlehem, PA, 18015}
\address[xt]{Department of Mathematics, University of California, San Diego, CA, 92093}
\address[xy]{Department of Industrial and Systems Engineering, Lehigh University, Bethlehem, PA, 18015}
\address[xl]{Department of Mathematics and Statistics, University of North Carolina at Charlotte, Charlotte, NC, 28223}
\address[np]{1 Science Center Drive, Corning Incorporated, Corning, NY, 14831}

\author[yy]{Yiming Fan}\ead{yif319@lehigh.edu}
\author[yy]{Huaiqian You}\ead{huy316@lehigh.edu}
\author[xt]{Xiaochuan Tian}\ead{xctian@ucsd.edu}
\author[xy]{Xiu Yang}\ead{xiy518@lehigh.edu}
\author[xl]{Xingjie Li}\ead{xli47@uncc.edu}
\author[np]{Naveen Prakash}\ead{PrakashN2@corning.com}
\author[yy]{Yue Yu}\ead{yuy214@lehigh.edu}


\begin{abstract}
In this work we aim to develop a unified mathematical framework and a reliable computational approach to model the brittle fracture in heterogeneous materials with variability in material microstructures, and to provide statistic metrics for quantities of interest, such as the fracture toughness. To depict the material responses and naturally describe the nucleation and growth of fractures, we consider the peridynamics model.
In particular, a stochastic state-based peridynamic model is developed, where the micromechanical parameters are modeled by a finite-dimensional random vector, or a combination of random variables truncating the Karhunen-Lo\`{e}ve decomposition or the principle component analysis (PCA). 
To solve this stochastic peridynamic problem, probabilistic collocation method (PCM) is employed to sample the random field representing the micromechanical parameters. For each sample, the deterministic peridynamic problem is discretized with an optimization-based meshfree quadrature rule. We present rigorous analysis for the proposed scheme and demonstrate its convergence for a number of benchmark problems, showing that it sustains the asymptotic compatibility spatially and achieves an algebraic or sub-exponential convergence rate in the random space as the number of collocation points grows. 
Finally, to validate the applicability of this approach on real-world fracture problems, we consider the problem of crystallization toughening in glass-ceramic materials, in which the material at the microstructural scale contains both amorphous glass and crystalline phases. The proposed stochastic peridynamic solver is employed to capture the crack initiation and growth for glass-ceramics with different crystal volume fractions, and the averaged fracture toughness are calculated. The numerical estimates of fracture toughness show good consistency with experimental measurements.
\end{abstract}


\end{frontmatter}


\tableofcontents

\section{Introduction}

Prediction and monitoring heterogeneous material damage are ubiquitous in applications of interest to the broad scientific and engineering community \cite{zohdi2002toughening,wriggers1998computational,prudencio2013dynamic,su2006guided,AFOSR2014,talreja2015modeling,soric2018multiscale,pijaudier2013damage,mourlas2019accurate,markou2021new}. In disciplines ranging from material design to non-destructive evaluation, heterogeneities in materials and media need to be accurately captured to guarantee reliable and trustworthy damage predictions that inform decision making. In the past decades, important discoveries and advancements have been made toward understanding material microstructures and its relationship with damage observed in the macroscale. New experimental technologies and test procedures have been designed to observe much smaller microstructure patterns and find defects in less time \cite{lindgren2013state,hdbk2009nondestructive,achenbach2000quantitative,forsyth2010air,aflcmc2013ez,jones2015probing,pan2018review,shukla2020physics}. On the other hand, novel mathematical models and numerical tools have been developed to describe failure initiation and progression, which provide relatively inexpensive alternatives to extensive experimental testing \cite{kok2018anisotropy,zhang2019review,bessa2017framework,bostanabad2018computational,han2020efficient}. However, fundamental challenges are still present in utilizing multiscale material models, and numerical simulations to provide a comprehensive physical and functional description of material damage, mainly due to the following difficulties \cite{lindgren2016us}:

\begin{enumerate}
\item The high degrees of complexity and heterogeneity in material damage problems generally require numerical simulations at fine scales that are often computationally prohibitive. For instance, bottom-up approaches such as the fine-grained atomistic models have provided important insights into the fracture process, but they generally do not scale up to finite-size samples. This limitation raises the need for new mathematical models that act at coarser scales and capture complex nonlinear modes of failure from the fine scale.
\item Different material microstructure, property, interfacial conditions, and operating environments all cause variability within material, which is tremendously difficult to be fully quantified. Therefore, without complete detailed measurements for each individual material sample, it is often non-practical, if not impossible, to provide full quantitative damage characterization for each sample. This fact calls for stochastic modeling of the variability and characterization of material failure for uncertainty quantification.
\end{enumerate}

These two challenges both call for mathematical models that not only capture the material fracture initiation and progression, but also account for heterogeneity and variability. To describe crack initiation and evolution simultaneously from the microscale-up, we employ the peridynamic theory, a spatially nonlocal continuum theory which provides a description of continuum mechanics in terms of integral operators rather than classical differential operators \cite{silling_2000,seleson2009peridynamics,parks2008implementing,zimmermann2005continuum,emmrich2007analysis,du2011mathematical,bobaru2016handbook,yu2018partitioned,trask2019asymptotically,yu2021asymptotically,you2022data,tian2013analysis,du2018peridynamic,prakash2016electromechanical,prakash2017computational,prakash2019calibrating}. These nonlocal models are defined in terms of a lengthscale $\delta$, referred to as a horizon, which denotes the range of nonlocal interaction between particles. The integral operator allows a natural description of processes requiring reduced regularity in the relevant solution, such as fracture mechanics \cite{bazant2002nonlocal,du2013nonlocal}. Therefore in peridynamics the material damage can be captured autonomously as a natural component of the material deformation. To account for heterogeneity and variability, we propose to develop a stochastic peridynamics formulation where the heterogeneous material property is modeled by a random field. Most of the current {state-of-the-art} works on peridynamics consider a homogenized and/or deterministic model, which may not work well when the material is heterogeneous and its microsctructure plays a critical role. In a recent study on reinforced concrete modeling, Zhao et al. found that a fully homogenized peridynamic model fails to capture certain correct fracture modes/patterns \cite{zhao2020stochastic}. Therefore, they have proposed a stochastic bond-based peridynamic model where the material property is described as random fields. The type of each bond connecting two material points $\xb$ and $\yb$ was modeled by a random variable, and the discrete probability distribution of this random variable depends on the volume fraction of aggregate and cement. With this model, fracture patterns match experimental observations. Their findings indicate the importance of considering the spatial variability of material properties in peridynamics. However, in \cite{zhao2020stochastic} the authors focused on the crack pattern in individual realizations rather than the solution statistics. Their numerical study only provides a qualitative validation on the fracture patterns and the order in which various cracks develop. To provide any quantitative verification and validation of the model, it calls for an effective stochastic method to provide the statistic metric on the impact of microstructure variability.

To this end, in this current work we propose a stochastic state-based peridynamics model where the heterogeneous material property is varying spatially and described by a random field. The solution of this stochastic problem describes the statistics of the material responses, such as the displacement and damage fields. In particular, we employ the linear peridynamic solid (LPS) model \cite{emmrich2007well} as a prototypical state-based model appropriate for brittle fracture, and propose a heterogeneous LPS formulation where two-point function formulations are used to describe the heterogeneous material properties. Although such an averaged two-point function formulation were developed for nonlocal diffusion \cite{fan2021asymptotically,guan2017reduced} and peridynamics \cite{nguyen2021depth,oterkus2014peridynamic,wang2015studies,behera2021peridynamic,mehrmashhadi2018effect} models, we have for the first time provided rigorous mathematical analysis for the well-posedness of this formulation in a heterogeneous LPS model. Furthermore, an important feature of peridynamics is that when classical continuum models still apply, peridynamics revert back to classical continuum models as its horizon size $\delta \rightarrow 0$. Numerical discretizations which preserve this limit under the grid refinement $h \rightarrow 0$ are termed asymptotically compatible (AC) \cite{tian2014asymptotically}, and there has been significant works in recent years toward establishing such discretizations \cite{tian2014asymptotically,d2020numerical,leng2019asymptotically,pasetto2018reproducing,hillman2020generalized,seleson2016convergence,du2016local,trask2019asymptotically,You_2019,you2020asymptotically,tao2017nonlocal,fan2021asymptotically}. In this work, we have also theoretically shown that our stochastic heterogeneous LPS model guarantees consistency to the corresponding local limit, which provides a critical ingredient in achieving a convergent simulation.

To enable numerical simulations to investigate the impact of microstructure variability, our second aim is to numerically discretize the proposed stochastic peridynamics model and provide the first two statistical moments, i.e., the mean and (co)variance. The mean provides an unbiased estimate of the variables and the variance quantifies the uncertainty associated with this estimate. Such a development calls for a comprehensive treatment of an AC spatial discretization method together with an effective stochastic method, which is able to perform convergent and efficient heterogeneous peridynamic fracture simulations while providing stochastic modeling of the variability and characterization of material failure for uncertainty quantification. Broadly, AC spatial discretization strategies for peridynamics can be classified into two categories. The first class involves traditional finite element formulations and carefully performing geometric calculations to integrate over relevant horizon/element subdomains, while the second type adopts a strong-form meshfree discretization where particles are associated with an abstract measure, and provides a sharp representation of the fracture surface by breaking bonds. The former is based on a variational setting and therefore is more amenable to mathematical analysis, while the latter is simple to implement and generally faster \cite{silling2005meshfree,bessa2014meshfree}. In this paper we pursue the meshfree viewpoint. In particular, a meshfree method is developed based on the optimization-based quadrature rule\footnote{For peridynamics one often refines both $\delta$ and $h$ at the same rate under so-called $\delta$-convergence \cite{bobaru2009convergence}. In this setting, banded stiffness matrices is obtained which allows scalable implementations. Although in the literature a scheme is termed AC if it recovers the solution whenever $\delta,h\rightarrow 0$, in this work we adopt a practical setting and only require the $\delta$-convergence case for AC.} \cite{trask2019asymptotically,yu2021asymptotically,fan2021asymptotically,foss2021convergence}. For the stochastic numerical method, several approaches were developed for stochastic local (classical) PDE models, including probabilistic Galerkin methods (PGMs) \cite{babuska2004galerkin,babuvska2005solving,ghanem2003stochastic,le2004uncertainty,matthies2005galerkin,xiu2002wiener,wan2005adaptive}, probablistic collocation methods (PCMs) \cite{xiu2005high,nobile2008anisotropic,ma2009adaptive,zhang2012error,lin2009efficient}, reduced basis methods \cite{rozza2007reduced,rozza2007stability,chen2014comparison,chen2013weighted,elman2013reduced,guan2017reduced}, etc. Among these methods, PCM with sparse grids inherits the ease of implementation in the Monte Carlo methods since only solutions at sample points are needed. At the same time, it also reduces the required number of sample points to achieve a given numerical accuracy for problems with relatively high dimension in the random space. Therefore, in this work we will employ PCM with full tensor products for random dimensions $N\leq 4$, and PCM with sparse grids when the dimension in the random dimension is larger than $4$, following the suggestion by \cite{lin2009efficient}. To verify and validate the proposed model and the numerical approach, we numerically investigate the convergence to the analytical local limit for a number of benchmark problems, including manufactured smooth solutions, composite material with discontinuous material properties, and material fracture problems. Last but not least, we validate estimates of fracture toughness on randomly heterogeneous materials against an experiment of glass-ceramics \cite{serbena2015crystallization}, providing evidence that the scheme yields accurate predictions for statistic damage metrics in practical engineering problems.

The paper is organized as follows. 
We describe first the deterministic and stochastic heterogeneous LPS problems in Section \ref{sec:math}, and provide mathematical analysis to establish their compatibility with the corresponding local problem. Next, we pursue a consistent discretization, and our numerical approach for stochastic LPS problems is proposed in Section \ref{sec:num} and numerically verified in Section \ref{sec:test}. When no fracture occurs and the material properties are sufficiently smooth, the classical continuum theory applies and the formulation preserves the AC limit under $\delta$-convergence, with an optimal $O(\delta^2)$ convergence rate. When fracture occurs and/or the material properties present discontinuity, the spatial discretization formulation is able to capture the material heterogeneity and the resultant damage field, with an $O(\delta)$ convergence rate to the local limit. When the nonlocal solution is analytic with respect to the input random variables, this method guarantees an at least algebraic convergence (for PCM with sparse grids) or exponential convergence (for PCM with full tensor products) with increasing sample numbers. Therefore, we have establish a unified mathematical framework, which is able to incorporate all of the necessary ingredients to perform non-trivial simulations of fracture mechanics in heterogeneous materials while maintaining a scalable implementation and guaranteeing convergence. In Section \ref{sec:val}, we further extend the proposed formulation to handle a more engineering-oriented problem, where a glassy matrix contains randomly distributed crystal grains. A quasi-static brittle fracture model is considered, to provide preliminary quantitative validation results by comparing our numerical results with available experimental measurements on material fracture toughness. Section \ref{sec:conclusion} summarizes our findings and discusses future research. Additional discussions and proofs for the truncation estimates between the local and nonlocal operators are provided in \ref{app:1}.

\section{Peridynamics for Randomly Heterogeneous Materials}\label{sec:math}

In this section, {we introduce the state-based peridynamics formulation}, together with the major notations and definitions. In particular, we will consider the linear peridynamic solid (LPS) model \cite{emmrich2007well}, which is a prototypical state-based model appropriate for brittle fracture. The LPS model may be interpreted as a nonlocal generalization of the mixed form of linear elasticity, evolving both displacements and a dilatation. We begin with a review of the deterministic LPS model for heterogeneous materials \cite{yu2021asymptotically} in Section \ref{sec:peri}, then extend the formulation to the stochastic LPS problem with random parameters in Section \ref{sec:randomperi}. Finally, we discuss the treatment of material fracture, including the damage criteria and the handling of free surfaces created by evolving fracture, in Section \ref{sec:neumann}.

\subsection{Deterministic Peridynamics Problem with Heterogeneous Material Properties}\label{sec:peri}

We begin by reviewing the governing equations of deterministic LPS models which provide the foundation for the stochastic problems of interest. In this section, we consider the material without damage, with fully prescribed Dirichlet type boundary conditions, and will further extend the discussions to more general boundary conditions and brittle fractures in Section \ref{sec:neumann}.

Consider a body occupying a bounded Lipschitz and convex domain $\Omega\subset \mathbb{R}^d$, $d = 2$ or $3$, with Dirichlet-type boundary conditions. Let $\ub:\omg\rightarrow \real^d$ be the displacement field, $\theta:\omg\rightarrow \real$ be the nonlocal dilatation, generalizing the local divergence of displacement, and $K:\real^d\times\real^d\rightarrow \real^+\cup\{0\}$ is a nonnegative kernel function. In this paper we further assume that the interacting kernel function $K$ is radially symmetric (which can therefore be denoted as $K(r)$ for $r\in\real^+\cup\{0\}$, with a slight abuse of notation), with compactly support on $B_\delta(0)$, the $\delta$-ball  centered at $0$, and satisfies the following conditions:
\begin{equation}\label{eqn:require_ga}
\left\{\begin{array}{l}
K(\xb,\yb)=K(|\xb-\yb|)=K_\delta(|\xb-\yb|)=\frac{1}{\delta^{d+2}}K_1\left(\frac{|\xb-\yb|}{\delta}\right),\\
\text{ where $K_1$ is nonnegative and there exists a positive constant $\zeta<1$ satisfying}\\
B_\zeta(\bm{0})\subset\text{supp}(K_1)\subset B_1(\bm{0}) \text{ and }\int_{B_1(\bm{0})}K_1(|\zb|)|\zb|^2d\zb=d.
\end{array}\right.
\end{equation}
The above kernel assumptions have implications on the boundary conditions that are prescribed on a collar of thickness $\delta$ near the boundary $\partial\Omega$, that we denote as
\begin{align*}
\omgi&:=\{\xb\in\Omega|\text{dist}(\xb,\partial\Omega)<\delta\},\,\omgb:=\{\xb\notin\Omega|\text{dist}(\xb,\partial\Omega)<\delta\},\,\omgbb:=\{\xb\notin\Omega|\text{dist}(\xb,\partial\Omega)<2\delta\}.
\end{align*}
To apply the nonlocal Dirichlet-type boundary condition, we assume that $\ub(\xb)=\ub_D(\xb)$ are provided in $\omgbb$.  Without loss of generality, for the analysis, we consider homogeneous Dirichlet boundary conditions $\ub_D(\xb)=\mathbf{0}$.

In the original LPS model for materials with homogeneous material properties \cite{emmrich2007well}, the momentum balance is given by the following
\begin{equation}\label{eq:nonlocElasticity}
\begin{aligned}
    \mcL_\delta \ub:=&-\frac{C_\alpha}{d}  \int_{B_\delta (\xb)} \left(\lambda- \mu\right) K(\left|\yb-\xb\right|) \left(\yb-\xb \right)\left(\theta(\xb) + \theta(\yb) \right) d\yb\\
  &\quad  
  -  \frac{C_\beta}{d}\int_{B_\delta (\xb)} \mu K(\left|\yb-\xb\right|)\frac{\left(\yb-\xb\right)\otimes\left(\yb-\xb\right)}{\left|\yb-\xb\right|^2}  \left(\ub(\yb) - \ub(\xb) \right) d\yb = \fb(\xb),\quad \text{ for }\xb\in\omg,
   \end{aligned}
\end{equation}
where the nonlocal dilatation $\theta(\xb)$ is defined as
\begin{equation}\label{eqn:oritheta}
\theta(\xb):=\int_{B_\delta (\xb)} K(\left|\yb-\xb\right|) (\yb-\xb)\cdot \left(\ub(\yb) - \ub(\xb) \right)d\yb,\text{ for }\xb\in\omg\cup\omgb.
\end{equation}
Here, $\fb\in\mathbb{R}^d$ denotes the external body loading forces, 
and $\mu$, $\lambda$ denote the (constant) shear modulus and Lam{\'{e}}  first parameter, respectively. With appropriate choice of scaling parameters $C_\alpha>0$, $C_\beta>0$ and the kernel function $K(r)$, it can be shown that the system converges to the Navier equations \cite{mengesha2012nonlocal,mengesha2014bond,mengesha2014nonlocal} for linear elasticity:
\begin{equation}\label{eqn:local}
\mcL_0 \ub:=-\nabla\cdot(\lambda \text{tr}(\mathbf{E})\mathbf{I}+2\mu \mathbf{E})=-(\lambda-\mu)\nabla [\text{tr}(\mathbf{E})]-\mu \nabla\cdot(2\mathbf{E}+\text{tr}(\mathbf{E})\mathbf{I})=\fb,
\end{equation}
where the strain tensor $\mathbf{E}:=\dfrac{1}{2}(\nabla \ub+(\nabla \ub)^T)$ and $\text{tr}(\mathbf{E})$ denotes its trace. To recover parameters for 3D linear elasticity, one should take $C_\alpha=3$, $C_\beta=30$; whereas for 2D problems, $C_\alpha=2$, $C_\beta=16$. In this paper we consider 2D problems ($d=2$), 
although the algorithm may be generalized to more general kernels and 3D cases.

In \cite{yu2021asymptotically}, the authors extended the above original LPS model to composite materials constituted of multiple phases, where the domain was partitioned into disjoint subdomains with piecewise constant material properties such that $\lambda(\xb)$ and $\mu(\xb)$ may vary for each material point $\xb$. In this work, we propose to further extend the original LPS model \eqref{eq:nonlocElasticity} and \eqref{eqn:oritheta} to the general heterogeneous materials, with either continuous or discontinuous material parameters $\lambda(\xb)$ and $\mu(\xb)$. Specifically, for the deterministic problem where the Lam{\'{e}} {moduli} $\lambda(\xb)$ and $\mu(\xb)$ may vary for each material point $\xb$, satisfying 
$$0<\lambda_0=\inf_{\xb\in{\omg\cup\omgbb}}\lambda(\xb)\leq \sup_{\xb\in{\omg\cup\omgbb}}\lambda(\xb)=\lambda_{\infty}<\infty,$$
$$0<\mu_0=\inf_{\xb\in{\omg\cup\omgbb}}\mu(\xb)\leq \sup_{\xb\in{\omg\cup\omgbb}}\mu(\xb)=\mu_{\infty}<\infty,$$
we employ the following momentum balance and nonlocal dilatation formulations: 
\begin{equation}\label{eq:nonlocElasticity_comp}
\begin{aligned}
    \mcL_{H\delta} \ub:=&-\int_{B_\delta (\xb)}  \left(\lambda(\xb,\yb) - \mu(\xb,\yb)\right)K(\left|\yb-\xb\right|) \left(\yb-\xb \right)\left(\theta(\xb) + \theta(\yb) \right) d\yb\\
 &\qquad
  -  8\int_{B_\delta (\xb)} \mu(\xb,\yb) K(\left|\yb-\xb\right|)\frac{\left(\yb-\xb\right)\otimes\left(\yb-\xb\right)}{\left|\yb-\xb\right|^2}  \left(\ub(\yb) - \ub(\xb) \right) d\yb = \fb(\xb),
  \end{aligned}
\end{equation}
where $\theta$ is defined in \eqref{eqn:oritheta}, and the two-point functions $\mu(\cdot,\cdot)$, $\lambda(\cdot,\cdot)$ denote averaged material properties. Specifically, we consider the interaction between $\xb$ and $\yb$ as a series of two springs connecting the two points, and then the equivalent total spring constant will be the harmonic mean of the two spring constants \cite{nguyen2021depth,mehrmashhadi2018effect,prakash2022investigation}: 
\begin{equation}\label{eqn:harmonic}
   \frac{2}{ \mu(\xb,\yb)} =\frac{1}{\mu(\xb)} + \frac{1}{\mu(\yb)},\quad\frac{2}{ \lambda(\xb,\yb)} =\frac{1}{\lambda(\xb)} + \frac{1}{\lambda(\yb)}.
\end{equation}
We notice that $\mu(\cdot,\cdot)$ and $\lambda(\cdot,\cdot)$ will also satisfy
$$0<\lambda_0=\inf_{\xb,\yb\in{\omg\cup\omgbb}}\lambda(\xb,\yb)\leq \sup_{\xb,\yb\in{\omg\cup\omgbb}}\lambda(\xb,\yb)=\lambda_{\infty}<\infty,$$
$$0<\mu_0=\inf_{\xb,\yb\in{\omg\cup\omgbb}}\mu(\xb,\yb)\leq \sup_{\xb,\yb\in{\omg\cup\omgbb}}\mu(\xb,\yb)=\mu_{\infty}<\infty.$$
For the proof of the algorithm's wellposedness, we will also need the following assumptions on $\lambda_0$, $\mu_0$, $\lambda_\infty$ and $\mu_\infty$:
\begin{assumption}\label{asp}
There exist two constants $A_0,A_1>0$ such that
\begin{align}
&(4-A_1)\mu_0-A_0(\lambda_\infty-\lambda_0)>0,\\
&\lambda_\infty-\dfrac{\lambda_\infty-\lambda_0}{2A_0}-\dfrac{\mu_\infty}{2A_1}\geq0.
\end{align}
\end{assumption}

\begin{remark}
We note that the above assumption generally requires a upper bound of $\mu_\infty$ and relatively small fluctuation of $\lambda(\xb)$ and $\mu(\xb)$. When considering homogeneous materials where the Lam\'{e} and shear modulus $\lambda(\xb)$ and $\mu(\xb)$ are both constants, we have $\lambda_\infty-\lambda_0=0$ and the two conditions yield $8\lambda_\infty>\mu_\infty$. This condition is suboptimal, since the homogeneous LPS model can be proved to be well-posed given any $\lambda,\mu>0$ (see, e.g., \cite{mengesha2014nonlocal}).

In Sections \ref{sec:test}-\ref{sec:val}, empirical experiments are performed on cases not satisfying Assumption \ref{asp}. Stable and converging numerical results are still observed. We will investigate more general and optimal conditions of $\lambda$ and $\mu$ in future work.
\end{remark}

Consider a (quasi) static state-based peridynamic problem with Dirichlet-type boundary conditions:
 \begin{equation}\label{eqn:probn}
\left\{\begin{array}{ll}
\mcL_{H\delta}\ub(\xb) = \fb(\xb),&\quad \text{ in }\omg\\
\theta(\xb)=\int_{B_\delta (\xb)} K(\left|\yb-\xb\right|) (\yb-\xb)\cdot \left(\ub(\yb) - \ub(\xb) \right)d\yb,&\quad \text{ in }\omg\cup\omgb\\
\ub(\xb)=\ub_D(\xb), &\quad \text{ in }\omgbb
\end{array}\right.
\end{equation}
multiply a test function $\vb(\xb)$ satisfying $\vb(\xb)=\mathbf{0}$ in $\omgbb$ to  \eqref{eq:nonlocElasticity_comp}, and integrate it with respect to $\xb\in {\omg\cup \omgbb} $, we then obtain the weak formulation
\begin{align*}
  (\fb,\vb&)_{L^2(\omg)}\\=&-  \int_{\omg\cup \omgbb} \int_{\omg\cup \omgbb}  \left(\lambda(\xb,\yb) - \mu(\xb,\yb)\right)K(\left|\yb-\xb\right|) \left(\yb-\xb \right)\cdot\vb(\xb)\left(\theta(\xb) + \theta(\yb) \right) d\yb d\xb\\
  &-  8\int_{\omg\cup \omgbb} \int_{\omg\cup \omgbb} \mu(\xb,\yb) K(\left|\yb-\xb\right|)\frac{\left(\yb-\xb\right)\otimes\left(\yb-\xb\right)}{\left|\yb-\xb\right|^2}  \left(\ub(\yb) - \ub(\xb) \right) d\yb \vb(\xb) d\xb\\
  =&  \int_{\omg\cup \omgbb} \int_{\omg\cup \omgbb}  \left(\lambda(\xb,\yb) - \mu(\xb,\yb)\right)K(\left|\yb-\xb\right|) \left(\yb-\xb \right)\cdot\left(\vb(\yb)-\vb(\xb)\right)d\yb\, \theta(\xb)d\xb\\
  &+  4\int_{\omg\cup \omgbb} \int_{\omg\cup \omgbb} \mu(\xb,\yb) \dfrac{K(\left|\yb-\xb\right|)}{\left|\yb-\xb\right|^2}[\left(\yb-\xb\right)\cdot\left(\ub(\yb) - \ub(\xb) \right)] [\left(\yb-\xb\right)\cdot\left(\vb(\yb) - \vb(\xb) \right)] d\yb d\xb\\
  :=&T_{H\delta}[\ub,\vb;\lambda,\mu],
\end{align*}
where $\theta$ is defined by \eqref{eqn:oritheta} and $\lambda$, $\mu$ are the two point material property functions defined in \eqref{eqn:harmonic}. And we denote the strain energy density function at material point $\xb$ as
\begin{align*}
W_{\ub}(\xb):=&\int_{\omg\cup \omgbb}  \left(\lambda(\xb,\yb) - \mu(\xb,\yb)\right)K(\left|\yb-\xb\right|) \left(\yb-\xb \right)\cdot\left(\vb(\yb)-\vb(\xb)\right)d\yb\, \theta(\xb)\\
  &+  4 \int_{\omg\cup \omgbb} \mu(\xb,\yb) \dfrac{K(\left|\yb-\xb\right|)}{\left|\yb-\xb\right|^2}[\left(\yb-\xb\right)\cdot\left(\ub(\yb) - \ub(\xb) \right)]^2 d\yb.    
\end{align*}

With the boundedness properties of $\lambda(\cdot,\cdot)$ and $\mu(\cdot,\cdot)$, we have the following charaterization of the space
\begin{lemma}\label{lemma:1}
The nonlocal energy semi-norm is
$$|\ub|_{S_{H\delta}(\omg)}= \left[\int_{\omg\cup \omgbb}  \int_{\omg\cup \omgbb} \dfrac{K(\left|\yb-\xb\right|)}{\left|\yb-\xb\right|^2}[\left(\yb-\xb\right)\cdot\left(\ub(\yb) - \ub(\xb) \right)]^2 d\bm y d\bm x\right]^{1/2},$$
and the nonlocal energy space is
$$S_{H\delta}(\omg)=\left\{\ub\in L^2(\omg;\real^d):\int_{\omg\cup\omgbb}  \int_{\omg\cup \omgbb} \dfrac{K(\left|\yb-\xb\right|)}{\left|\yb-\xb\right|^2}[\left(\yb-\xb\right)\cdot\left(\ub(\yb) - \ub(\xb) \right)]^2 d\bm y d\bm x<\infty,\,\ub|_{\omgbb}=\mathbf{0}\right\}.$$
\end{lemma}
\begin{proof}
With the Cauchy-Schwartz inequality we have
\begin{align}
|\theta(\xb)|&=
\verti{\int_{\omg\cup \omgbb} K(\left|\yb-\xb\right|) (\yb-\xb)\cdot \left(\ub(\yb) - \ub(\xb) \right)d\yb}\nonumber\\
&\leq\sqrt{2}\left(\int_{\omg\cup \omgbb} \dfrac{K(\left|\yb-\xb\right|)}{\left|\yb-\xb\right|^2}[\left(\yb-\xb\right)\cdot\left(\ub(\yb) - \ub(\xb) \right)]^2d\yb \right)^{1/2},\label{eqn:fortheta1}
\end{align}
and 
\begin{align}
\nonumber&\int_{\omg\cup \omgbb}\int_{\omg\cup \omgbb}  K(\left|\yb-\xb\right|) \verti{\left(\yb-\xb \right)\cdot\left(\ub(\yb)-\ub(\xb)\right)}d\yb\, \theta(\xb)d\xb\\
\nonumber\leq&\sqrt{2}\int_{\omg\cup \omgbb}\left(\int_{\omg\cup \omgbb}  \dfrac{K(\left|\yb-\xb\right|)}{\left|\yb-\xb\right|^2}[\left(\yb-\xb\right)\cdot\left(\ub(\yb) - \ub(\xb) \right)]^2d\yb \right)^{1/2}|\theta(\xb)|d\xb\\
\leq& \dfrac{1}{2\tilde{A}}\int_{\omg\cup \omgbb} (\theta(\xb))^2 d\xb+ \tilde{A}\int_{\omg\cup \omgbb}\int_{\omg\cup \omgbb}  \dfrac{K(\left|\yb-\xb\right|)}{\left|\yb-\xb\right|^2}[\left(\yb-\xb\right)\cdot\left(\ub(\yb) - \ub(\xb) \right)]^2d\yb d\xb,\label{eqn:fortheta2}
\end{align}
where the second inequality comes from the Young's inequality with any positive constant $\tilde{A}$. We then insert \eqref{eqn:fortheta1} into \eqref{eqn:fortheta2} to get
\begin{equation}\label{eqn:fortheta3} 
\begin{split}
\int_{\omg\cup \omgbb}&\int_{\omg\cup \omgbb}  K(\left|\yb-\xb\right|) \verti{\left(\yb-\xb \right)\cdot\left(\ub(\yb)-\ub(\xb)\right)}d\yb\, |\theta(\xb)|\, d\xb\\
&\le \dfrac{1}{2\tilde{A}}\int_{\omg\cup \omgbb} (\theta(\xb))^2 d\xb+ \tilde{A}\int_{\omg\cup \omgbb}\int_{\omg\cup \omgbb}  \dfrac{K(\left|\yb-\xb\right|)}{\left|\yb-\xb\right|^2}[\left(\yb-\xb\right)\cdot\left(\ub(\yb) - \ub(\xb) \right)]^2d\yb d\xb\\
&\le\left(\frac{1}{\tilde{A}}+\tilde{A}\right)\int_{\omg\cup \omgbb}\int_{\omg\cup \omgbb}  \dfrac{K(\left|\yb-\xb\right|)}{\left|\yb-\xb\right|^2}[\left(\yb-\xb\right)\cdot\left(\ub(\yb) - \ub(\xb) \right)]^2d\yb d\xb.
    \end{split}
\end{equation}
\noindent 
We then prove that any $\ub\in S_{H\delta}(\omg)$ has a bounded total strain energy. Taking $\tilde{A}=1$, for any $\ub \in S_{H\delta}(\omg)$ its total strain energy satisfies 
\begin{align*}
\int_{\omg\cup \omgbb}W_{\ub}(\xb)d\xb\leq&  \int_{\omg\cup \omgbb} \int_{\omg\cup \omgbb}  (\lambda(\xb,\yb)+\mu(\xb,\yb))K(\left|\yb-\xb\right|) \verti{\left(\yb-\xb \right)\cdot\left(\ub(\yb)-\ub(\xb)\right)}d\yb\, |\theta(\xb)|d\xb\\
  &+  4\int_{\omg\cup \omgbb} \int_{\omg\cup \omgbb} \mu(\xb,\yb) \dfrac{K(\left|\yb-\xb\right|)}{\left|\yb-\xb\right|^2}[\left(\yb-\xb\right)\cdot\left(\ub(\yb) - \ub(\xb) \right)]^2 d\yb d\xb\\ 
\leq&(2 \lambda_{\infty}+6\mu_{\infty})\int_{\omg\cup \omgbb} \int_{\omg\cup \omgbb} \dfrac{K(\left|\yb-\xb\right|)}{\left|\yb-\xb\right|^2}[\left(\yb-\xb\right)\cdot\left(\ub(\yb) - \ub(\xb) \right)]^2 d\yb d\xb<\infty.
\end{align*}
On the other hand, for any $\ub$ satisfying $\int_{\omg\cup \omgbb}W_{\ub}(\xb)d\xb\leq\infty$, we aim to show that $\ub\in S_{H\delta}(\omg)$. In particular,
\begin{align*}
\int_{\omg\cup \omgbb}W_{\ub}(\xb)d\xb=&  \int_{\omg\cup \omgbb} \int_{\omg\cup \omgbb}  (\lambda(\xb,\yb)-\mu(\xb,\yb))K(\left|\yb-\xb\right|) {\left(\yb-\xb \right)\cdot\left(\ub(\yb)-\ub(\xb)\right)}d\yb\, \theta(\xb) d\xb\\
  &+  4\int_{\omg\cup \omgbb} \int_{\omg\cup \omgbb} \mu(\xb,\yb) \dfrac{K(\left|\yb-\xb\right|)}{\left|\yb-\xb\right|^2}[\left(\yb-\xb\right)\cdot\left(\ub(\yb) - \ub(\xb) \right)]^2 d\yb d\xb\\
\geq&\int_{\omg\cup \omgbb} \int_{\omg\cup \omgbb}  \lambda(\xb,\yb)K(\left|\yb-\xb\right|) {\left(\yb-\xb \right)\cdot\left(\ub(\yb)-\ub(\xb)\right)}d\yb\, \theta(\xb) d\xb\\
  &+  A_1\int_{\omg\cup \omgbb} \int_{\omg\cup \omgbb} \mu(\xb,\yb) K(\left|\yb-\xb\right|)\left[\dfrac{\left(\yb-\xb\right)\cdot\left(\ub(\yb) - \ub(\xb) \right)}{\left|\yb-\xb\right|}-\dfrac{1}{2A_1}\left|\yb-\xb\right|\theta(\xb)\right]^2 d\yb d\xb\\
&+(4-A_1)\int_{\omg\cup \omgbb} \int_{\omg\cup \omgbb}\mu(\xb,\yb) \dfrac{K(\left|\yb-\xb\right|)}{\left|\yb-\xb\right|^2}[\left(\yb-\xb\right)\cdot\left(\ub(\yb) - \ub(\xb) \right)]^2 d\yb d\xb\\  
&-  \frac{1}{4A_1}\int_{\omg\cup \omgbb} \int_{\omg\cup \omgbb} \mu(\xb,\yb) K(\left|\yb-\xb\right|)\left|\yb-\xb\right|^2 d\yb (\theta(\xb))^2 d\xb
\end{align*}
{{Since for $d=2$ and \eqref{eqn:require_ga} we have}}
\begin{align*}
&- \int_{\omg\cup \omgbb} \int_{\omg\cup \omgbb} \mu(\xb,\yb) K(\left|\yb-\xb\right|)\left|\yb-\xb\right|^2 d\yb (\theta(\xb))^2 d\xb\\
&\quad \geq- \mu_{\infty}\int_{\omg\cup \omgbb} \int_{\omg\cup \omgbb} K(\left|\yb-\xb\right|)\left|\yb-\xb\right|^2 d\yb (\theta(\xb))^2 d\xb=-2\mu_{\infty}\int_{\omg\cup \omgbb}  (\theta(\xb))^2 d\xb,
\end{align*}
and by taking $\tilde{A}=A_0(\lambda_{\infty}-\lambda_0)$ in \eqref{eqn:fortheta2}
\begin{align*}
 &  \int_{\omg\cup \omgbb} \int_{\omg\cup \omgbb}  \lambda(\xb,\yb)K(\left|\yb-\xb\right|) [{\left(\yb-\xb \right)\cdot\left(\ub(\yb)-\ub(\xb)\right)}]d\yb\, \theta(\xb) d\xb\\
 =&  \int_{\omg\cup \omgbb} \int_{\omg\cup \omgbb}  (\lambda(\xb,\yb)-\lambda_\infty) K(\left|\yb-\xb\right|) [{\left(\yb-\xb \right)\cdot\left(\ub(\yb)-\ub(\xb)\right)}]d\yb\, \theta(\xb) d\xb+\lambda_\infty\int_{\omg\cup \omgbb} (\theta(\xb))^2 d\xb\\
 \geq&\lambda_\infty\int_{\omg\cup \omgbb} (\theta(\xb))^2 d\xb-(\lambda_\infty-\lambda_0)  \int_{\omg\cup \omgbb} \int_{\omg\cup \omgbb}  K(\left|\yb-\xb\right|) \verti{\left(\yb-\xb \right)\cdot\left(\ub(\yb)-\ub(\xb)\right)}d\yb\, |\theta(\xb)| d\xb\\
 \geq 
 &\left(\lambda_\infty-\dfrac{(\lambda_\infty-\lambda_0)}{2A_0} \right)\int_{\omg\cup \omgbb} (\theta(\xb))^2 d\xb- A_0(\lambda_\infty-\lambda_0)\int_{\omg\cup \omgbb}\int_{\omg\cup \omgbb}  \dfrac{K(\left|\yb-\xb\right|)}{\left|\yb-\xb\right|^2}[\left(\yb-\xb\right)\cdot\left(\ub(\yb) - \ub(\xb) \right)]^2d\yb d\xb,
\end{align*}
substituting the above two inequalities yields:
\begin{align*}
\int_{\omg\cup \omgbb}W_{\ub}(\xb)d\xb\geq&A_1\int_{\omg\cup \omgbb} \int_{\omg\cup \omgbb} \mu(\xb,\yb) K(\left|\yb-\xb\right|)\left[\dfrac{\left(\yb-\xb\right)\cdot\left(\ub(\yb) - \ub(\xb) \right)}{\left|\yb-\xb\right|}-\dfrac{1}{2A_1}\left|\yb-\xb\right|\theta(\xb)\right]^2 d\yb d\xb\\
&+((4-A_1)\mu_0-A_0(\lambda_\infty-\lambda_0))\int_{\omg\cup \omgbb} \int_{\omg\cup \omgbb} \dfrac{K(\left|\yb-\xb\right|)}{\left|\yb-\xb\right|^2}[\left(\yb-\xb\right)\cdot\left(\ub(\yb) - \ub(\xb) \right)]^2 d\yb d\xb\\
&+\left(\lambda_\infty-\dfrac{\lambda_\infty-\lambda_0}{2A_0}-\dfrac{\mu_\infty}{2A_1}\right)\int_{\omg\cup \omgbb} (\theta(\xb))^2 d\xb\\
\geq&((4-A_1)\mu_0-A_0(\lambda_\infty-\lambda_0))\int_{\omg\cup \omgbb} \int_{\omg\cup \omgbb} \dfrac{K(\left|\yb-\xb\right|)}{\left|\yb-\xb\right|^2}[\left(\yb-\xb\right)\cdot\left(\ub(\yb) - \ub(\xb) \right)]^2 d\yb d\xb.
\end{align*}
Therefore, $\int_{\omg\cup \omgbb} \int_{\omg\cup \omgbb} \dfrac{K(\left|\yb-\xb\right|)}{\left|\yb-\xb\right|^2}[\left(\yb-\xb\right)\cdot\left(\ub(\yb) - \ub(\xb) \right)]^2 d\yb d\xb<\infty$ and $\ub\in S_{H\delta}(\omg)$.
\end{proof}

\begin{remark}
Note that the above derivation also holds for the local extremes of $\lambda(\cdot,\cdot)$ and $\mu(\cdot,\cdot)$. Therefore an alternative (local) form of Assumption \ref{asp} writes:
\begin{align}
&(4-A_1)\mu_0(\xb)-A_0(\lambda_\infty(\xb)-\lambda_0(\xb))>0,\\
&\lambda_\infty(\xb)-\dfrac{\lambda_\infty(\xb)-\lambda_0(\xb)}{2A_0}-\dfrac{\mu_\infty(\xb)}{2A_1}\geq0,
\end{align}
for any $\xb\in\omg\cup\omgbb$, where 
$$\lambda_\infty(\xb):=\underset{\yb\in B_{\delta}(\xb)\cap(\omg\cup\omgbb)}{\sup}\lambda(\xb,\yb),\;\lambda_0(\xb):=\underset{\yb\in B_{\delta}(\xb)\cap(\omg\cup\omgbb)}{\inf}\lambda(\xb,\yb),$$
$$\mu_\infty(\xb):=\underset{\yb\in B_{\delta}(\xb)\cap(\omg\cup\omgbb)}{\sup}\mu(\xb,\yb),\;\mu_0(\xb):=\underset{\yb\in B_{\delta}(\xb)\cap(\omg\cup\omgbb)}{\inf}\mu(\xb,\yb).$$
If further assuming that $\lambda(\cdot),\mu(\cdot)\in C(\overline{\omg\cup\omgbb})$, we will have 
$$\underset{\xb\in\omg\cup\omgbb}{\sup}\;\underset{\yb,\zb\in B_{\delta}(\xb)\cap(\omg\cup\omgbb)}{\sup}|\lambda(\xb,\yb)-\lambda(\xb,\zb)|\leq C_1\delta,$$
$$\underset{\xb\in\omg\cup\omgbb}{\sup}\;\underset{\yb,\zb\in B_{\delta}(\xb)\cap(\omg\cup\omgbb)}{\sup}|\mu(\xb,\yb)-\mu(\xb,\zb)|\leq C_2\delta,$$
for generic constant $C_1$ and $C_2$ which are independent of $\delta$. Then Assumption \ref{asp} can be relaxed to:
\begin{align}
&(4-A_1)\mu_0(\xb)>A_0C_1\delta,\;\lambda_\infty(\xb)-\dfrac{\mu_\infty(\xb)}{2A_1}\geq\dfrac{C_2\delta}{2A_0}.
\end{align}
\end{remark}

In the rest of this paper, we will use $\vertii{\ub}_{L^2}$ to denote the $L^2(\omg;\real^d)$ norm of $\ub$, and $\vertii{\ub}_{S_{H\delta}}$ to denote the norm on $S_{H\delta}(\omg)$:
$$\vertii{\ub}^2_{S_{H\delta}}=\vertii{\ub}^2_{L^2}+\verti{\ub}^2_{S_{H\delta}}.$$
With the equivalance of the total strain energy with the $\verti{\cdot}_{S_{H\delta}}$ seminorm proved in Lemma \ref{lemma:1}, similar as in \cite{mengesha2014nonlocal} we have the following characterization of the zero energy solution:

\begin{lemma}
For all $\ub\in S_{H\delta}(\omg)$, $\int_{\omg\cup\omgbb} W_{\ub}(\xb)d\xb\geq0$, and 
$$\ub={\bf 0} {\text{ in }\omg\cup\omgbb } \Longleftrightarrow \verti{\ub}_{S_{H\delta}}=0 \Longleftrightarrow \int_{\omg\cup\omgbb} W_{\ub}(\xb)d\xb=0.$$
\end{lemma}

Following the proof of \cite[Proposition~2]{mengesha2014nonlocal}, we also have the nonlocal Poincare inequality:

\begin{lemma}\label{Lemma:nonlocal_Poincare}
Suppose that $V$ is a closed subspace of $L^2(\omg;\real^d)$, 
then there exists $C$ such that
$$\vertii{\ub}_{L^2}\leq C \int_{\omg\cup \omgbb} \int_{\omg\cup \omgbb} \dfrac{K(\left|\yb-\xb\right|)}{\left|\yb-\xb\right|^2}[\left(\yb-\xb\right)\cdot\left(\ub(\yb) - \ub(\xb) \right)]^2 d\yb d\xb,\;\forall \ub\in V.$$
Here $C$ is a generic constant depending on $K$, $V$ and $\omg$. Consequently, there exists a generic constant $\tilde{C}$ such that
$$\vertii{\ub}_{L^2}\leq {\tilde{C}} \verti{\ub}_{S_{H\delta}},\;\forall \ub\in S_{H\delta}(\omg).$$
\end{lemma}



\noindent
With the above lemmas, we obtain the coercivity and continuity of the bilinear form $T_{H\delta}[\ub,\vb;\lambda,\mu]$:

\begin{lemma}
There exist two constants $r, C >0$ such that 
\begin{align}
\text{Coercivity:}\quad&T_{H\delta}[\ub,\ub;\lambda,\mu]\geq r \verti{\ub}^2_{S_{H\delta}},\\
\text{Continuity:}\quad&T_{H\delta}[\ub,\vb;\lambda,\mu]\leq C \verti{\ub}_{S_{H\delta}}\verti{\vb}_{S_{H\delta}},
\end{align}
for any $\ub,\vb\in S_{H\delta}(\omg)$.
\end{lemma}
\begin{proof}
The coercivity of $T_{H\delta}$ is an immediate result of Lemma \ref{lemma:1}. The continuity is obtained by applying \eqref{eqn:fortheta1} and \eqref{eqn:fortheta2}.
\end{proof}

Finally, denoting the dual space of $S_{H\delta}(\omg)$ as $S_{H\delta}(\omg)^\ast$, the well-posedness result is obtained as an application of the Lax-Milgram theorem:

\begin{theorem}\label{thm:wellposed}
For a given body load $\fb\in S_{H\delta}(\omg)^\ast$, there exists a unique $\ub\in S_{H\delta}(\omg)$ such that
$$T_{H\delta}[\ub,\vb;\lambda,\mu]=\langle\fb,\vb\rangle,\;\forall \vb \in S_{H\delta}(\omg).$$
\end{theorem}

With the well-posedness proved, we now investigate the consistency of the proposed nonlocal formulation with the classical linear elastic model as $\delta \rightarrow 0$. Specifically, the classical linear elastic model with heterogeneous material parameters writes:\footnote{We note that it is generally not necessary to have the local solution $\ub$ defined in $\omgbb$. When $\omg$ is a Lipschitz domain, the above bounds can also be obtained for the general ${\ub}\in C^4(\overline{\omg})$, since one can extend $\ub$ to a $C^4$ function $\hat{\ub}$ in $\real^d$ (see, e.g., \cite[Section~2.5]{brudnyi2011methods}). For further discussions on applying Dirichlet-type boundary conditions as an extended local solution, we refer interested readers to \cite{foss2021convergence}.}
 \begin{equation}\label{eq:u0}
 \left\{\begin{array}{ll}
 \mathcal{L}_{H0}\ub(\xb):=-(\lambda(\xb)-\mu(\xb))\nabla [\text{tr}(\mathbf{E}(\xb))]-\mu(\xb) \nabla\cdot(2\mathbf{E}(\xb)+\text{tr}(\mathbf{E}(\xb))\mathbf{I})=\fb(\xb),&\, \text{in }\Omega,\\
 \ub{(\xb)}= \ub_D{(\xb)},&\, \text{on }\omgbb,
 \end{array}\right.
 \end{equation}
where $\mathbf{E}(\xb):=\dfrac{1}{2}(\nabla \ub(\xb)+(\nabla \ub(\xb))^T)$. We denote $\ub_\delta$ as the solution of the peridynamics problem \eqref{eqn:probn} and $\ub_0$ as the solution of \eqref{eq:u0}, and aim to show that $\ub_\delta\rightarrow\ub_0$ as $\delta\rightarrow0$. 

We first study the consistency of operators 
with the following lemma. 
Detailed proofs are elaborated in \ref{app:1}. 


\begin{lemma}\label{lem:truncation}
{Assume that $\ub\in C^4(\overline{\omg\cup\omgbb})$ and $\lambda(\cdot),\mu(\cdot)\in C^2(\overline{\omg\cup\omgbb})$, 
then there exists $\overline{\delta}>0$ such that for any $0<\delta\leq \overline{\delta}$,   $|\mcL_{H0}\ub(\xb)-\mcL_{H\delta} \ub(\xb)|\leq C\delta^2$ for $\xb\in \Omega$. Here the generic constant $C$ is independent of $\delta$ but may depend on the $C^4$ norm of $\ub$.}
\end{lemma}

With above regularity assumptions on $\ub_0$ and $\lambda(\cdot)$, $\mu(\cdot)$, we now further investigate the convergence of $\vertii{\ub_{\delta}-\ub_{0}}_{L^2}$:
\begin{theorem}
\label{thm:compatibility}
Let $\ub_\delta$ be the weak solution to the nonlocal problem and $\ub_0$ the weak solution to the local problem. 
Assume that 
{$\lambda(\cdot),\mu(\cdot)\in C(\overline{\omg\cup\omgbb})$. }
Then for any $\fb\in (S_{H \delta}(\omg))^\ast$, the dual space of $S_{H \delta}(\omg)$, we have 
\begin{equation}
\label{eqn:energy_estimate}
\| \ub_\delta\|_{S_{H \delta}(\omg)}\leq  \frac{ \| \fb\|_{(S_{H \delta}(\omg))^\ast}}{r}.   
\end{equation}
In addition,  if $\| \fb\|_{(S_{H \delta}(\omg))^\ast}$ is uniformly bounded for all $\delta\in (0,\delta_0)$ for some postive constant $\delta_0>0$, then the nonlocal and local diffusion problems are compatible as $\delta \rightarrow 0$:
$$\underset{\delta\rightarrow 0}{\lim}\vertii{\ub_\delta-\ub_0}_{L^2(\omg;\real^d)}=0.$$
\end{theorem}

\begin{proof}
We first show the proof of \eqref{eqn:energy_estimate}. Since $\ub_\delta$ is a solution to the nonlocal problem, we have
\[
T_{H \delta}[\ub_\delta, \vb;\lambda,\mu] = \langle \fb,\vb\rangle \leq \| \fb\|_{(S_{H \delta}(\omg))^\ast} \| \vb\|_{S_{H \delta}(\omg)}
\]
for any test function $\vb\in S_{H \delta}(\omg)$. Now let $\vb= \ub_\delta$, we get $r\| \ub_\delta\|^2_{S_{H \delta}(\omg)}\leq T_{H \delta}[\ub_\delta, \ub_\delta]  \leq  \| \fb\|_{(S_{H \delta}(\omg))^\ast} \|  \ub_\delta\|_{S_{H \delta}(\omg)}$. 
Therefore, we have \eqref{eqn:energy_estimate}. 

The proof of the second part involves two steps. In the first step, we assume $\lambda(\cdot), \mu(\cdot) \in C^2(\overline{\omg\cup\omgbb})$, 
then from Lemma \ref{lem:truncation}, we know that $ \mcL_{H\delta} \vb$ converges to $\mcL_0 \vb$ uniformly on $\omg$ for $\vb\in C_0^\infty(\omg)$ as $\delta\to 0$.  Notice that from the assumption on $\| \fb\|_{(S_{H \delta}(\omg))^\ast}$, we have  $\| \ub_\delta\|_{S_{H \delta}(\omg)}$ being uniformly bounded for all $\delta\in (0,\delta_0)$. Then using similar arguments in \cite{tian2014asymptotically} together with the compactness result \cite[Lemma 7]{mengesha2014nonlocal},  we can show $\|\ub_\delta -\ub_0 \|_{L^2(\omg;\real^d)}\to 0$ as $\delta\to0$.

 For the general case that $\lambda, \mu \in C(\overline{\omg\cup\omgbb})$, we will use the mollification technique.
 First notice that we can extend $\lambda$ and $\mu$ continuously to a larger domain that contains $\overline{\omg\cup\omgbb}$. 
 Then we can take standard mollifiers $\phi^\epsilon \in C^\infty(\R^{d})$, and define $\lambda^\epsilon = \phi^\epsilon \ast \lambda$ and $\mu^\epsilon = \phi^\epsilon \ast \mu $ on $\overline{\omg\cup\omgbb}$ for small enough $\epsilon>0$.  We denote the solution to \eqref{eqn:probn} associated with coefficient $\lambda^\epsilon(\xb, \yb):=  2((\lambda^\epsilon(\xb))^{-1} +  (\lambda^\epsilon(\yb))^{-1})^{-1} $ and $\mu^\epsilon(\xb, \yb):=  2((\mu^\epsilon(\xb))^{-1} +  (\mu^\epsilon(\yb))^{-1})^{-1} $ to be $\ub_{\delta, \epsilon}$. Then we can use the first step to conclude that 
$
\|\ub_{\delta, \epsilon} - \ub_{0, \epsilon} \|_{L^2(\omg;\real^d)} \xrightarrow{\delta\to0} 0
$, where $\ub_{0,\epsilon}$ is the solution to \eqref{eq:u0} associated with coefficient $\lambda^\epsilon$ and $\mu^\epsilon$. Now in order to show $\| \ub_\delta -\ub_0\|_{L^2(\omg;\real^d)}\to 0$, we notice that
\[
\lim_{\delta\to0}\| \ub_\delta - \ub_0\|_{L^2(\omg;\real^d)} \leq \sup_{\delta\in (0,\delta_0)} \| \ub_{\delta, \epsilon}-\ub_{\delta} \|_{L^2(\omg;\real^d)} +  \lim_{\delta\to0}\|\ub_{\delta, \epsilon} - \ub_{0, \epsilon} \|_{L^2(\omg;\real^d)} +  \| \ub_{0, \epsilon}-\ub_{0} \|_{L^2(\omg;\real^d)} ,
\]
for any $\epsilon>0$.
Therefore, we only need to show 
\begin{equation}\label{eqn:step2estimates}
\left\{ 
\begin{aligned}
& \lim_{\epsilon\to0}\sup_{\delta\in (0,\delta_0)} \| \ub_{\delta, \epsilon}-\ub_{\delta} \|_{L^2(\omg;\real^d)}=0, \quad \text{ and } \\
&\lim_{\epsilon\to0} \| \ub_{0, \epsilon}-\ub_{0} \|_{L^2(\omg;\real^d)}=0. 
\end{aligned} 
\right.
\end{equation}
Notice that $\| \lambda^\epsilon -\lambda\|_{C(\overline{\omg\cup\omgbb})} \to 0 $ and  $\| \mu^\epsilon -\mu\|_{C(\overline{\omg\cup\omgbb})} \to 0 $ as $\epsilon\to0$  since $\lambda$ and $\mu$ are continuous. Then $\lambda^\epsilon(\xb,\yb)\to \lambda(\xb,\yb)$ and $\mu^\epsilon(\xb,\yb)\to \mu(\xb,\yb)$  uniformly on $(\overline{\omg\cup\omgbb})^2$ as $\epsilon\to0$. For this, we simply write   $\| \lambda^\epsilon -\lambda\|_{C((\overline{\omg\cup\omgbb})^2)} \to 0 $ and  $\| \mu^\epsilon -\mu\|_{C((\overline{\omg\cup\omgbb})^2)} \to 0 $  where the functions $\lambda^\epsilon$, $\lambda$, $\mu^\epsilon$ and $\mu$ are continuous functions of the two variables $\xb\in\overline{\omg\cup\omgbb}$ and $\yb\in\overline{\omg\cup\omgbb}$. 
Now for the first equation in \eqref{eqn:step2estimates}, notice that since $\ub_{\delta, \epsilon}$ and $\ub_\delta$ are solutions to \eqref{eqn:probn} with different coefficients and the same right-hand side, we have 
\[
T_{H \delta}[\ub_{\delta, \epsilon} - \ub_{\delta}, \vb ; \lambda^\epsilon, \mu^\epsilon]  = T_{H \delta}[\ub_\delta, \vb ; \lambda-\lambda^\epsilon, \mu-\mu^\epsilon]  = : \langle \gb_{\delta,\epsilon}, \vb\rangle, 
\]
for any $\vb\in S_{H\delta}(\omg)$. We can show $ \langle \gb_{\delta,\epsilon}, \vb\rangle \to 0$ as $\epsilon\to0$ uniformly independent of $\delta$ since 
\[
\begin{split}
 &\langle \gb_{\delta,\epsilon}, \vb\rangle \\
 \leq & \left(\| \lambda^\epsilon -\lambda\|_{C((\overline{\omg\cup\omgbb})^2)} + \| \mu^\epsilon -\mu\|_{C((\overline{\omg\cup\omgbb})^2)}\right)\iint_{(\omg\cup \omgbb)^2} K(\left|\yb-\xb\right|) | \left(\yb-\xb \right)\cdot\left(\vb(\yb)-\vb(\xb)\right)|d\yb\, |\theta_\delta(\xb)|d\xb\\
  &+  4 \| \mu^\epsilon -\mu\|_{C((\overline{\omg\cup\omgbb})^2)} \iint_{(\omg\cup \omgbb)^2} \dfrac{K(\left|\yb-\xb\right|)}{\left|\yb-\xb\right|^2} | \left(\yb-\xb\right)\cdot\left(\ub_\delta(\yb) - \ub_\delta(\xb) \right)| |\left(\yb-\xb\right)\cdot\left(\vb(\yb) - \vb(\xb) \right)| d\yb d\xb  \\
 \leq &  5 \left(\| \lambda^\epsilon -\lambda\|_{C((\overline{\omg\cup\omgbb})^2)} + \| \mu^\epsilon -\mu\|_{C((\overline{\omg\cup\omgbb})^2)}\right)  \| \ub_\delta \|_{S_{H\delta}(\omg)} \| \vb \|_{S_{H\delta}(\omg)}. 
  \end{split}
  \] 
Now  use the coercivity of $T_{H\delta}$ and $ \| \ub_\delta \|_{S_{H \delta}(\omg)} \leq C$ from \eqref{eqn:energy_estimate}, we have
\[
\begin{split}
&\sup_{\delta\in (0,\delta_0)}\| \ub_{\delta, \epsilon}-\ub_{\delta} \|_{S_{H \delta}(\omg)} \leq C\sup_{\delta\in (0,\delta_0)} \| \gb_{\delta, \epsilon}\|_{(S_{H \delta}(\omg))^\ast} \\
\leq &C \left(\| \lambda^\epsilon -\lambda\|_{C((\overline{\omg\cup\omgbb})^2)} + \| \mu^\epsilon -\mu\|_{C((\overline{\omg\cup\omgbb})^2)}\right)  \to 0 \text{ as } \epsilon\to0
\end{split}
\]
 and the convergence in $L^2$ is then implied from the Poincar\'e inequality in Lemma~\ref{Lemma:nonlocal_Poincare}. 
The proof for the second equation in \eqref{eqn:step2estimates} can be similarly done. 
\end{proof}

The next theorem characterizes the rate of convergence of $\ub_\delta$ to $\ub_0$ as $\delta\to0$ when additional regularity is assumed for $\ub_0$. 
\begin{theorem}\label{thm:compatibility_with_rate}
Let $\ub_\delta$ be the weak solution to the nonlocal problem and $\ub_0$ the weak solution to the local problem. In addition, we assume that $\ub_0\in C^4(\overline{\omg\cup\omgbb})$ and $\lambda(\cdot),\mu(\cdot)\in C^2(\overline{\omg\cup\omgbb})$. Then there exists $\overline{\delta}>0$ such that for any $0<\delta\leq \overline{\delta}$, we have
\[
\| \ub_0 -\ub_\delta \|_{S_{H\delta}(\omg)} \leq C \delta^2, 
\]
where the generic constant $C$ is independent of $\delta$ but may depend on the $C^4$ norm of $\ub_0$.
\end{theorem}
\begin{proof}
Since $\ub_0$ is defined on $\overline{\omg\cup\omgbb}$, then we can compute $-\mcL_{H\delta} \ub_0(\xb) $ for any $\xb\in\omg$. Let $\fb_\delta(\xb)=-\mcL_{H\delta} \ub_0(\xb)$, then we have $ - \eb_\delta(\xb) = -\mcL_{H\delta}(\ub_\delta -\ub_0)(\xb)=\mcL_{H\delta}(\ub_0)(\xb)-\mcL_{H0}(\ub_0)(\xb) = \fb(\xb)-\fb_\delta(\xb)$ for $\xb\in\omg$. From the truncation error analysis in Lemma \ref{lem:truncation}, we get $\| \fb-\fb_\delta\|_{L^\infty(\omg,\mathbb{R}^d)}=O(\delta^2)$. Since $\ub_\delta -\ub_0$ is the weak solution to the nonlocal problem with load vector $\fb-\fb_\delta$, we can use \eqref{eqn:energy_estimate} to get
\[
\| \ub_0 -\ub_\delta \|_{S_{H\delta}(\omg)} \leq C \|f-f_\delta \|_{L^2(\omg,\mathbb{R}^d)} \leq C \|f-f_\delta \|_{L^\infty(\omg,\mathbb{R}^d)} = O(\delta^2).
\]
\end{proof}

\subsection{Parametric Peridynamics Problem}\label{sec:randomperi}

In this section, we will consider the case where the material properties $\lambda$ and $\mu$ are provided by random fields  $\lambda(\xb,\omega)$ and $\mu(\xb,\omega)$, where $\omega\in\Omega_p$ and $\Omega_p$ is the sample space of a probability space $(\Omega_p, \mathcal{F}, \mathcal{P})$. Here,
$\mathcal{F}$ is the $\sigma$-algebra of subsets of $\Omega_p$ and $\mathcal{P}$ is the probability measure. Following the practice in \cite{fan2021asymptotically}, we represent this random field in a ``truncated'' form using a limited number of random variables, either because they have been approximated by a truncated expansion such as the Karhunen-Loeve expansion or through PCA (see Section \ref{sec:val}), or because the input itself is defined in terms of a finite number of random variables. Thus, the material parameters can be rewritten as $\lambda(\xb, \xib)$ and $\mu(\xb, \xib)$, where $\xib=(\xi_{(1)},\xi_{(2)},\dotsc, \xi_{(N)})$, $N$ is a positive integer {which denotes the dimension of the parametric space}, and $\xi_{(i)}$ are random variables, and we assume they are independent and identically distributed (i.i.d.) random variables. Under this setting, we consider 
$$\lambda(\xb,\xib):(\omg\cup\omgbb)\times\Gamma\rightarrow \real,\quad\mu(\xb,\xib):(\omg\cup\omgbb)\times\Gamma\rightarrow \real,$$
where $\Gamma$ is the space of $\xib$ and it is typically called random space or parametric space. 
Without loss of generality, here we assume that $\Gamma= \prod_{i=1}^N \Gamma_{i} \subset \R^N$ where $\Gamma_{i} = [-1,1]$, and the random variable $\xib\in \Gamma$ has a probability density $\rho:\Gamma \to \R^+$. Similar as in the deterministic problem, for each $\xib\in\Gamma$, we use harmonic means of $\lambda$ and $\mu$ to model averaged material properties:
\begin{equation}\label{eqn:harmonic_xi}
   \frac{2}{ \mu(\xb,\yb,\xib)} =\frac{1}{\mu(\xb,\xib)} + \frac{1}{\mu(\yb,\xib)},\quad\frac{2}{ \lambda(\xb,\yb,\xib)} =\frac{1}{\lambda(\xb,\xib)} + \frac{1}{\lambda(\yb,\xib)}.
\end{equation}
We are then interested in solving the family of heterogeneous peridynamic problems given by 
\begin{equation}\label{eqn:nonlocal_random}
\left\{\begin{array}{ll}
\mcL_{H\delta} \ub:=-\int_{B_\delta (\xb)}  \left(\lambda(\xb,\yb,\xib) - \mu(\xb,\yb,\xib)\right)K(\left|\yb-\xb\right|) \left(\yb-\xb \right)\left(\theta(\xb,\xib) + \theta(\yb,\xib) \right) d\yb&\\
~~~~~ -  8\int_{B_\delta (\xb)} \mu(\xb,\yb,\xib) K(\left|\yb-\xb\right|)\frac{\left(\yb-\xb\right)\otimes\left(\yb-\xb\right)}{\left|\yb-\xb\right|^2}  \left(\ub(\yb,\xib) - \ub(\xb,\xib) \right) d\yb = \fb(\xb),\; &\text{ for }\xb\in\omg,\\
\theta(\xb,\xib)=\int_{B_\delta (\xb)}  K(\left|\yb-\xb\right|) (\yb-\xb)\cdot \left(\ub(\yb,\xib) - \ub(\xb,\xib) \right)d\yb,\; &\text{ for }\xb\in\omg\cup\omgb, \\
\ub(\xb,\xib)=\ub_D(\xb,\xib),\; &\text{ for }\xb\in\omgbb.
\end{array}\right.
\end{equation}
For each $\xib\in \Gamma$, 
we assume the uniform boundedness of the material properties, i.e., 
\begin{align*}\label{eqn:Uniform}
0<\lambda_0=\inf_{\xb\in{\omg\cup\omgbb}}\lambda(\xb,\xib)\leq \sup_{\xb\in{\omg\cup\omgbb}}\lambda(\xb,\xib)=\lambda_{\infty}<\infty,\\
0<\mu_0=\inf_{\xb\in{\omg\cup\omgbb}}\mu(\xb,\xib)\leq \sup_{\xb\in{\omg\cup\omgbb}}\mu(\xb,\xib)=\mu_{\infty}<\infty,
\end{align*}
for $\xib\in\Gamma$, and $\lambda_0$, $\lambda_{\infty}$, $\mu_0$, $\mu_\infty$ satisfy Assumption \ref{asp}. Therefore, for each $\xib\in\Gamma$, the conditions for Theorem \ref{thm:wellposed} still hold and therefore the Lax-Milgram theorem ensures the well-posedness of the corresponding peridynamic problem. 
In addition, in order to consider the limit $\delta\to0$, we need to assume that for each $\xib\in\Gamma$ and $\xb\in\omg\cup\omgbb$, 
\begin{equation}\label{eqn:DiagCont}
\lambda(\cdot, \xib),\mu(\cdot, \xib)\in C(\overline{\omg\cup\omgbb}). 
\end{equation}
Then we have the corresponding family of local linear elastic problem for each $\xib\in\Gamma$:
\begin{equation}\label{eqn:local_random}
 \left\{\begin{array}{ll}
 \mathcal{L}_{H0}\ub(\xb, \xib):=-(\lambda(\xb, \xib)-\mu(\xb, \xib))\nabla [\text{tr}(\mathbf{E}(\xb, \xib))]-\mu(\xb, \xib) \nabla\cdot(2\mathbf{E}(\xb, \xib)+\text{tr}(\mathbf{E}(\xb, \xib))\mathbf{I})=\fb(\xb),&\, \text{in }\Omega,\\
 \ub(\xb,\xib)= \ub_D(\xb,\xib),&\, \text{in }\omgbb.
 \end{array}\right.
 \end{equation}

For each given parameter $\xib \in \Gamma$, we denote the solution to the peridynamic problem \eqref{eqn:nonlocal_random} by $\ub_\delta(\xb, \xib)$ and the solution to the corresponding local equation \eqref{eqn:local_random} by  $\ub_0(\xb, \xib)$. A corollary of Theorem \ref{thm:compatibility} is that $\ub_\delta(\xb, \xib)$ converges to $\ub_0(\xb, \xib)$ in the space $L^2(\omg)\otimes L^2_{\rho}(
\Gamma)$ as $\delta\to0$:


\begin{corollary}\label{thm:compatible}
Let $\ub_\delta(\xb,\xib)$ be the weak solution to \eqref{eqn:nonlocal_random} and $\ub_0(\xb,\xib)$ the weak solution to \eqref{eqn:local_random}. Assume that $\lambda(\cdot,\xib), \mu(\cdot,\xib) \in C(\overline{\omg\cup\omgbb})$, then there exists $\overline{\delta}>0$ such that for any $0<\delta\leq \overline{\delta}$, we have 
{\[
\lim_{\delta\rightarrow 0}\| \ub_\delta - \ub_0\|_{L^2( \omg;\real^d)\otimes  L^2_\rho(\Gamma)}= 0\,.
\]}
In addition, if we have $\ub_0(\cdot,\xib)\in C^4(\overline{\omg\cup\omgbb})$ with uniform $C^4$ norm for $\xib\in \Gamma$ and $\lambda(\cdot,\xib),\mu(\cdot,\xib)\in C^2(\overline{\omg\cup\omgbb})$, then 
\[
\| \ub_\delta - \ub_0\|_{S_{H\delta}(\omg)\otimes  L^2_\rho(\Gamma)} \leq C \delta^2,
\]
{where the generic constant $C$ is independent of $\delta$ but may depend on the $C^4$ norm of $\ub_0$.}
\end{corollary}
\begin{proof}
For any $\xib\in \Gamma$, and, we know from Theorem \ref{thm:compatibility} that  $\| \ub_\delta(\cdot, \xib)\|_{S_{H \delta}(\omg)}\leq C $ for all $\delta\in (0,\delta_0)$ and  $\| \ub_\delta(\cdot, \xib) - \ub_0(\cdot, \xib)\|_{L^2(\omg;\real^d)}\to 0$ as $\delta\to0$. Therefore, it is easy to see that $\| \ub_\delta(\cdot, \xib) - \ub_0(\cdot, \xib)\|_{L^2(\omg;\real^d)} \leq C$ for all $\xib\in\Gamma$ and $\delta\in (0,\delta_0)$.  Using the dominated convergence theorem, we have
\[
\| \ub_\delta - \ub_0 \|_{ L^2(\omg;\real^d) \otimes L^2_\rho(\Gamma) } 
= \int_{\Gamma} \| \ub_\delta(\cdot , \xib)- \ub_0( \cdot, \xib) \|^2_{L^2(\omg;\real^d)}   \rho(\xib)d\xib \overset{\delta\rightarrow 0}{\longrightarrow} 0 \,.
\]
The second statement comes from Theorem \ref{thm:compatibility_with_rate} by noticing that $\| \ub_0(\cdot,\xib)\|_{C^4(\overline{\omg\cup\omgbb})}\leq C$ for all $\xib\in \Gamma$. 
\end{proof}

\subsection{Peridynamics Formulation for Brittle Fractures}\label{sec:neumann}

One of the main appeals of peridynamics is to handle fracture problems, where free surfaces are associated with the evolution of a fracture surface. In this section, we first consider the deterministic LPS model and propose the handling of free surfaces in heterogeneous materials, then apply it to the treatment of material fracture. Lastly, we will conclude this section with a stochastic LPS formulation for evolving fracture.

We now consider general mixed boundary conditions: $\partial\Omega=\partial\Omega_D\bigcup \partial\Omega_N$ %
and $(\partial\Omega_D)^o\bigcap (\partial\Omega_N)^o=\emptyset$. Here $\partial \Omega_D$ and $\partial \Omega_N$ are both curves. To apply the nonlocal Dirichlet-type boundary condition, we assume that $\ub(\xb)=\ub_D(\xb)$ are provided in a layer with non-zero volume outside $\Omega$, while the free surface boundary condition is applied on the sharp interface $\partial\Omega_N$. To define a Dirichlet-type constraint, we denote
\begin{align*}
\omgi_D&:=\{\xb\in\Omega|\text{dist}(\xb,\partial\Omega_D)<\delta\},\,\omgb_D:=\{\xb\notin\Omega|\text{dist}(\xb,\partial\Omega_D)<\delta\},\,\omgbb_D:=\{\xb\notin\Omega|\text{dist}(\xb,\partial\Omega_D)<2\delta\},
\end{align*}
and assume that the value of $\ub$ is given on $\omgbb_D$. For notation simplicity, we denote $\omg_D:=\omg\cup\omgbb_D$. Similarly, to apply the free surface boundary condition, we denote
\begin{align*}
\omgi_N&:=\{\xb\in\Omega|\text{dist}(\xb,\partial\Omega_N)<\delta\},\,\omgb_N:=\{\xb\notin\Omega|\text{dist}(\xb,\partial\Omega_N)<\delta\},\,\omgbb_N:=\{\xb\notin\Omega|\text{dist}(\xb,\partial\Omega_N)<2\delta\}.
\end{align*}
Unless stated otherwise, in this paper we further assume sufficient regularity in the boundary that we may take $\delta$ sufficiently small so that for any $\xb\in \omgi$, there exists a unique orthogonal projection\footnote{{Here we notice that it is possible $\omgi_D\cap\omgi_N\neq\emptyset$. In our numerical solver, we treat $\xb$ with the Dirichlet-type boundary condition if the projection of $\xb$ is in $\partial\omg_D$. Otherwise, we use the Neumann-type boundary condition at $\xb$.}} of $\xb$ onto $\partial\Omega$, which is the closest point on $\partial\Omega$ to $\xb$. We denote this projection as $\overline{\xb}$. Therefore, one has $\overline{\xb}-\xb=s_x\mathbf{n}({\xb})$ for $\xb\in \omgi_N$, where $0<s_x<\delta$. Here $\mathbf{n}$ denotes the normal direction pointing out of the domain for each $\xb\in\omgi_N$, and let $\mathbf{p}$ denote the tangential direction. 
Here, we propose the following formulation for the (partially) free surface problem:
 \begin{align}
    \nonumber\mcL_{N\delta}\ub(\xb):=&-\int_{B_\delta (\xb)\cap\omg_D} \left(\lambda(\xb,\yb) - \mu(\xb,\yb)\right) K(\left|\yb-\xb\right|) 
     \left(\yb-\xb \right)\left(\theta^{corr}(\xb) + \theta^{corr}(\yb) \right) d\yb\\
   \nonumber&-8\int_{B_\delta (\xb)\cap\omg_D}\mu(\xb,\yb)
     K(\left|\yb-\xb\right|)\frac{\left(\yb-\xb\right)\otimes\left(\yb-\xb\right)}{\left|\yb-\xb\right|^2} 
      \left(\ub(\yb) - \ub(\xb) \right) d\yb\\
    \nonumber&-2\theta^{corr}(\xb) \int_{B_\delta (\xb)\backslash\omg_D}
     \left(\lambda(\xb,\yb) - \mu(\xb,\yb)\right) K(\left|\yb-\xb\right|) 
     \left(\yb-\xb \right) d\yb\\
\nonumber&-4\theta^{corr}(\xb)\int_{B_\delta (\xb)\backslash\omg_D}(\lambda(\xb,\yb)+2\mu(\xb,\yb)) K(\left|\yb-\xb\right|)
     \frac{[\left(\yb-\xb \right)\cdot \mathbf{n}][\left(\yb-\xb \right)\cdot \mathbf{p}]^2}{\left|\yb-\xb\right|^2}
      \mathbf{n}d\yb \\
&+4\theta^{corr}(\xb) \int_{B_\delta (\xb)\backslash\omg_D} \lambda(\xb,\yb) K(\left|\yb-\xb\right|)   \frac{[\left(\yb-\xb \right)\cdot \mathbf{n}]^3}{\left|\yb-\xb\right|^2}\mathbf{n} d\yb= \fb(\xb),\label{eq:newform1}
 \end{align}
and
\begin{equation}\label{eq:continuousNonlocdilatation3_new}
  \theta^{corr}(\xb) = \int_{B_\delta (\xb)\cap\omg_D} K(\left|\yb-\xb\right|) \left(\yb-\xb\right) \cdot \mathbf{M}(\xb)\cdot \left(\ub(\yb) - \ub(\xb) \right) d\yb,
\end{equation}
\begin{equation}\label{eq:continuousdilCorr_new}
  \mathbf{M}(\xb) =  \left[ \int_{B_\delta (\xb)\cap\omg_D}K(\left|\yb-\xb\right|) \left(\yb-\xb\right) \otimes \left(\yb-\xb\right)  d\yb \right]^{-1}.
\end{equation}
Here we notice that for $\xb\notin\omgi_N$, 
$\mathbf{M}(\xb)$ coincides with the identity matrix and hence $\theta^{corr}=\theta$. Therefore, the nonlocal operator $\mcL_{N\delta}$ in \eqref{eq:newform1} is the same as $\mcL_{H\delta}$ for $\xb\in\omg\backslash\omgi_N$. That means, for material points which are sufficiently far away from the free surface, we obtain the momentum balance and nonlocal dilatation formulation \eqref{eq:nonlocElasticity_comp}. On the other hand, when considering homogeneous materials, i.e., when $\lambda(\xb)=\lambda$ and $\mu(\xb)=\mu$ are constants, we obtain the Neumann-type LPS formulation developed in \cite{yu2021asymptotically}, which as shown to provide an approximation for the corresponding linear elastic model with free surfaces in the case of linear displacement fields.


With the free surface formulation, 
we now employ the composite LPS model \eqref{eq:nonlocElasticity_comp} and extend it to model brittle fracture in the general heterogeneous materials. 
In peridynamics, material damage is incorporated into the constitutive model by allowing the bonds of material points to break irreversibly. To model brittle fracture in the LPS model, we employ the critical stretch criterion where breakage occurs when a bond is extended beyond some predetermined critical bond deformed length \cite{zhang2018state,yu2021asymptotically}. Although a similar idea can be applied for dynamic fracture problems \cite{yu2021asymptotically}, in this work we consider quasi-static fracture problems, and use the time instant $t$ to denote the indexes for (incrementally increasing) loading in quasi-static problems. For example, the displacement solution at time instant $t$ will be denoted as $\ub(\xb,t)$. 
Consider the case where the material properties $\lambda$, $\mu$ and the fracture energy $G$ are provided by random fields  $\lambda(\xb,\xib)$, $\mu(\xb,\xib)$ and $G(\xb,\xib)$, where we recall that $\xib=(\xi_{(1)},\xi_{(2)},\dotsc, \xi_{(N)})$, with $N\in\mathbb{N}$ being {the dimension of the parametric space}, and $\xi_{(i)}$ are i.i.d. random variables. We propose the following formulation for $\xb\in\omg$
 \begin{align}
    \nonumber\mcL_{F\delta}&\ub(\xb,t,\xib):=-\int_{B_\delta(\xb)} \gamma(\xb,\yb,t,\xib)\left(\lambda(\xb,\yb,\xib) - \mu(\xb,\yb,\xib)\right) K(\left|\yb-\xb\right|) 
     \left(\yb-\xb \right)\left(\theta^{corr}(\xb,t,\xib) + \theta^{corr}(\yb,t,\xib) \right) d\yb\\
   \nonumber&-8\int_{B_\delta (\xb)} \gamma(\xb,\yb,t,\xib)\mu(\xb,\yb,\xib)
     K(\left|\yb-\xb\right|)\frac{\left(\yb-\xb\right)\otimes\left(\yb-\xb\right)}{\left|\yb-\xb\right|^2} 
      \left(\ub(\yb,t,\xib) - \ub(\xb,t,\xib) \right) d\yb\\
    \nonumber&-2\theta^{corr}(\xb,t,\xib)\int_{B_\delta (\xb)}(1-\gamma(\xb,\yb,t,\xib)) 
     \left(\lambda(\xb,\yb,\xib) - \mu(\xb,\yb,\xib)\right) K(\left|\yb-\xb\right|) 
     \left(\yb-\xb \right) d\yb\\
\nonumber&-4\theta^{corr}(\xb,t,\xib)\int_{B_\delta (\xb)}(1-\gamma(\xb,\yb,t,\xib)) (\lambda(\xb,\yb,\xib)+2\mu(\xb,\yb,\xib)) K(\left|\yb-\xb\right|)
     \frac{[\left(\yb-\xb \right)\cdot \mathbf{n}][\left(\yb-\xb \right)\cdot \mathbf{p}]^2}{\left|\yb-\xb\right|^2}
      \mathbf{n}d\yb \\
&+4\theta^{corr}(\xb,t,\xib) \int_{B_\delta (\xb)} (1-\gamma(\xb,\yb,t,\xib))\lambda(\xb,\yb,\xib) K(\left|\yb-\xb\right|)   \frac{[\left(\yb-\xb \right)\cdot \mathbf{n}]^3}{\left|\yb-\xb\right|^2}\mathbf{n} d\yb= \fb(\xb,t),\label{eq:newform2}
 \end{align}
and for $\xb\in\omg\cup\omgb_D$
\begin{equation}\label{eq:continuousNonlocdilatation3_new_f}
  \theta^{corr}(\xb,t,\xib) = \int_{B_\delta (\xb)}\gamma(\xb,\yb,t,\xib) K(\left|\yb-\xb\right|) \left(\yb-\xb\right) \cdot \mathbf{M}(\xb,t,\xib)\cdot \left(\ub(\yb,t,\xib) - \ub(\xb,t,\xib) \right) d\yb,
\end{equation}
\begin{equation}\label{eq:continuousdilCorr_new_f}
  \mathbf{M}(\xb,t,\xib) =  \left[ \int_{B_\delta (\xb)}\gamma(\xb,\yb,t,\xib) K(\left|\yb-\xb\right|) \left(\yb-\xb\right) \otimes \left(\yb-\xb\right)  d\yb \right]^{-1},
\end{equation}
where the averaged two-point functions $\mu(\cdot,\cdot,\xib)$, $\lambda(\cdot,\cdot,\xib)$ are defined using the harmonic mean, following \eqref{eqn:harmonic_xi}. The boolean state function $\gamma(\xb,\yb,t,\xib)$ is defined and updated following 
\begin{align}\label{eqn:gamma}
    \gamma(\xb,\yb,t,\xib) &= \begin{cases}
    1, \quad \text{if }s(\xb,\yb,\tau,\xib)\leq s_0(\xb,\yb,\xib),\;\forall\tau\leq t, \text{ and }\yb\in B_\delta(\xb)\cap\Omega_D,\\
    0, \quad \text{otherwise}, \\
    \end{cases}
\end{align}
with the associated strain $s$ and the critical bond stretch $s_0$ related to material parameters:
\begin{align}
         \nonumber&s(\xb,\yb,t,\xib): = \frac{||\ub(\yb,t,\xib)-\ub(\xb,t,\xib) + \yb-\xb||-||\yb - \xb||}{||\yb - \xb||},\\
         &s_0(\xb,\yb,\xib) := \sqrt{\frac{G(\xb,\yb)}{4(\lambda(\xb,\yb,\xib) - \mu(\xb,\yb,\xib))\beta' + 8\mu(\xb,\yb,\xib) \beta}},\,\text{ where }\beta := \frac{3\delta}{4\pi},\,\beta' := 0.23873 \delta.\label{eq:criteria}
\end{align}
Here $G(\xb,\yb,\xib)$ is the averaged fracture energy defined via the arithmetic mean:
\begin{equation}\label{def_fract_energy}
   G(\xb,\yb,\xib)= \frac{1}{2}(G(\xb,\xib)+G(\yb,\xib)).
\end{equation}
To summarize, for each $\xib\in\Gamma$, we obtain a unified mathematical formulation for a (quasi)-static state-based peridynamic problem with general mixed boundary conditions for brittle fractures:
 \begin{equation}\label{eqn:probn_rand}
\left\{\begin{array}{ll}
\mcL_{F\delta}\ub(\xb,t,\xib) = \fb(\xb,t),&\quad \text{ in }\omg\\
\theta^{corr}(\xb,t,\xib)=\int_{B_\delta (\xb)} \gamma(\xb,\yb,t,\xib)K(\left|\yb-\xb\right|) (\yb-\xb)^T \mathbf{M}(\xb,t,\xib)\left(\ub(\yb,t,\xib) - \ub(\xb,t,\xib) \right)d\yb,&\quad \text{ in }\omg\cup\omgb_D\\
\ub(\xb,t,\xib)=\ub_D(\xb,t,\xib), &\quad \text{ in }\omgbb_D
\end{array}\right.
\end{equation}

\begin{remark}
To see the intuition for the averaged material properties definition in \eqref{eqn:harmonic_xi} and the averaged fracture energy definition in \eqref{def_fract_energy}, we take the interaction between $\xb$ and $\yb$ as an analog of a series of two springs connecting the two points. Assuming that the two springs are with elongation lengths $l_1$ and $l_2$, respectively, and their spring constants are $k_1$  and $k_2$, respectively. We notice that $l_1$ and $l_2$ can be seen as the analog of the bond elongation in peridynamics, i.e., $\ub(\yb)-\ub((\xb+\yb)/2)$ and $\ub((\xb+\yb)/2)-\ub(\xb)$, respectively, and $k_1$, $k_2$ can be seen as the analog of material properties. Then the force balance between  $\xb$, $\yb$ yields $k_1l_1=k_2l_2$ and therefore the equivalent strength of this bond would be $k=\dfrac{k_1l_1+k_2l_2}{l_1+l_2}=\dfrac{2}{k_1^{-1}+k_2^{-1}}$, which can be viewed as a simplified version of the harmonic mean formulation for the averaged material properties definition in \eqref{eqn:harmonic_xi}. On the other hand, the total energy of the spring series writes $\dfrac{1}{2}(k_1l_1^2+k_2l_2^2)$, hence we define the averaged fracture energy via the arithmetic mean, as shown in \eqref{def_fract_energy}\footnote{We note that in some studies the harmonic mean formulation is employed for the averaged fracture energy (see \cite{nguyen2021depth} and references therein), which would make the interfacial bonds relatively weaker than what we proposed here. However, as studied in \cite{agwai2011predicting}, in bimaterial problems the interfacial bond strength depends on the interfacial adhesion strength, which should be provided by experiments. Therefore, without further measurements from experiments, we employ the arithmetic mean definition here since it provides a better agreement of fracture toughness with experimental measurements in Section \ref{sec:val}.}.
\end{remark}

\section{Spatial and Stochastic Numerical Methods}\label{sec:num}

In this section, we firstly introduce a strong form of meshfree discretization for the stochastic LPS model. Specifically, the optimization-based quadrature rule \cite{yu2021asymptotically,trask2019asymptotically,fan2021asymptotically} will be employed for spatial discretization, which is simple to implement and generally faster \cite{silling2005meshfree,bessa2014meshfree}, and was shown to be asymptotically compatible with corresponding local solutions in the absence of fracture \cite{yu2021asymptotically}. To sample the random field, the probabilistic collocation method (PCM) is employed, for its high accuracy and ease of implementation by sampling at discrete points in a random space \cite{tatang1994direct,keese2003numerical,xiu2005high}. Of course, the main appeal of peridynamic discretizations is to handle fracture problems. Therefore, we will also demonstrate how the meshfree scheme adapts to the brittle fracture formulation described in Section \ref{sec:neumann}, where free surfaces are associated with the time evolution of a fracture surface. Finally, the fully-discretized formulation for heterogeneous LPS model with random microstructure will be considered. In absence of fracture and assuming that the solution possesses sufficient continuity, we show that the proposed formulation sustains the asymptotic compatibility spatially and achieves an algebraic or sub-exponential convergence rate in the random coefficients space as the number of collocation points grows. When fracture occurs, our formulation automatically provides a sharp representation of the fracture surface by breaking bonds for each microstructure, and then estimates of quantities of interest in heterogeneous material damage problems, such as the fracture toughness, can be obtained.

\subsection{Spatial: Optimization-Based Meshfree Quadrature Rules}\label{sec:meshfree}

Discretizing the whole interaction region $\Omega\cup\omgbb$ by a collection of points $\chi_{h} = \{\xb_i\}_{\{i=1,2,\cdots,M\}} \subset \Omega\cup\omgbb$, we aim to solve for the displacement $u_{i}\approx u(\xb_i)$ and nonlocal dilitation $\theta_{i}\approx \theta(\xb_i)$ on each $\xb_i\in \chi_h$. Recall the definitions \cite{wendland2004scattered} of fill distance $h_{\chi_h,\Omega} = \underset{\xb_i \in \Omega\cup\omgbb}{\sup}\, \underset{\xb_j \in \chi_h}{\min}||\xb_i - \xb_j||_2$ and separation distance
${q_{\chi_h} = \frac12 \underset{i \neq j}{\min} ||\xb_i - \xb_j||_2}$. 
For simplicity we drop subscripts and simply write $h$ and $q$. In this paper we assume that $\chi_h$ is \textit{quasi-uniform}, namely that there exists $c_{qu} > 0$ such that $q_{\chi_h} \leq h_{\chi_h,\Omega} \leq c_{qu} q_{\chi_h}$. To maintain an easily scalable implementation, we further assume $\delta$ to be chosen such that the ratio $\frac{h}{\delta}$ is bounded as $\delta \rightarrow 0$, restricting ourselves to the ``$\delta$-convergence'' scenario \cite{bobaru2009convergence}.


Following \cite{yu2021asymptotically}, for materials without fracture we then pursue a discretization in 2D space of the system \eqref{eq:nonlocElasticity} and \eqref{eqn:oritheta} through the following one point quadrature rule at $\chi_h$ \cite{silling2010peridynamic}:
\begin{align}%
    (\mathcal{L}_{H\delta}^h \ub)_i:=&\sum_{\xb_j \in \chi_h\cap B_\delta(\xb_i)}  \left(\lambda_{ij} - \mu_{ij}\right)K_{ij} \left(\xb_j-\xb_i \right)\left(\theta_i + \theta_j \right) \omega_{j,i}\label{eq:nonlocElasticity_num}\\
  \nonumber&+ 8\sum_{\xb_j \in \chi_h\cap B_\delta(\xb_i)} \mu_{ij} K_{ij}\frac{\left(\xb_j-\xb_i\right)\otimes\left(\xb_j-\xb_i\right)}{\left|\xb_j-\xb_i\right|^2}  \left(\ub_i - \ub_j \right) \omega_{j,i} = \fb_i,\\
\theta_i:=&\sum_{\xb_j \in \chi_h\cap B_\delta(\xb_i)}  K_{ij} (\xb_j-\xb_i)\cdot \left(\ub_j - \ub_i \right) \omega_{j,i},\label{eqn:theta_num}
\end{align}
where we adopt notations $q_i=q(\xb_i), q_{ij}=q(\xb_i,\xb_j)$ for generic functions $q$. $\{\omega_{j,i}\}_{\xb_j \in B_\delta (\xb_i)}$ is a collection of to-be-determined quadrature weights corresponding to a neighborhood of collocation point $\xb_i$, which will be constructed through an optimization-based approach in \cite{fan2021asymptotically} to ensure consistency guarantees. Specifically, we seek quadrature weights for integrals supported on balls of the form
\begin{equation}
I[q] := \int_{B_\delta (\xb_i)} q(\xb_i,\yb) d\yb \approx I_h[q] := \sum_{\xb_j \in \chi_h\cap B_\delta(\xb_i)\backslash\{\xb_i\}} q(\xb_i,\xb_j)\omega_{j,i}
\end{equation}
where the subscript $i$ in $\left\{\omega_{j,i}\right\}$ denote that we seek a different family of quadrature weights for different subdomains $B_\delta(\xb_i)$. These weights are then generated from the following optimization problem
\begin{align}\label{eq:quadQP}
  \underset{\left\{\omega_{j,i}\right\}}{\text{argmin}} \sum_{\xb_j \in \chi_h\cap B_\delta(\xb_i)\backslash\{\xb_i\}} \omega_{j,i}^2 \quad
  \text{such that}, \quad
  I_h[q] = I[q] \quad \forall q \in \bm{V}_{h,\xb_i},
\end{align}
where  $\bm{V}_{h,\xb_i}=\left\{q(\yb-\xb_i)=\frac{p(\yb-\xb_i)}{|\yb-\xb_i|^3}\Big|p\in\mathbb{P}_5(\mathbb{R}^d)\text{ such that } \int_{B_\delta(\xb_i)}q(\yb-\xb_i)d\yb<\infty\right\}$ denotes the space of functions which should be integrated exactly. $\mathbb{P}_m(\real^d)$ is the space of $m$-th order polynomials. 
As shown in \cite{yu2021asymptotically}, for $\ub_0 \in C^{4}(\overline{\omg\cup\omgbb})$ this particular choice of reproducing space guarantees that the truncation error for all nonlocal operators in \eqref{eq:nonlocElasticity} converge to its local limit with an $O(\delta^2)$ rate in the limit $\delta \rightarrow 0$. For further discussions and error estimates of this optimization-based quadrature rule, we refer interested readers to \cite{fan2021asymptotically}.


\subsection{Stochastic:  Probabilistic Collocation Method with Sparse Grids}

In this work, we use the probabilistic collocation method (PCM) in the parametric space to solve the parametric peridynamics problem \cite{tatang1994direct,keese2003numerical,xiu2005high}. Consider the stochastic LPS Problem \eqref{eqn:nonlocal_random}, PCM can be seen as a Lagrange interpolation in the random space. In particular, let $\Theta_N=\{\xib_k\}_{k=1}^Q\subset \Gamma$ be a set of prescribed nodes such that the Lagrange interpolation in the random space $\Gamma$ is poised in an interpolation space $\Gamma_I$, where $N$ is the dimension of the parametric space. Then any function $v:\Gamma\rightarrow \real$ can be approximated using the Lagrange interpolation polynomial $\mathcal{J}[v](\xib)=\sum_{k=1}^Q v(\xib_k)J_k(\xib)$, where $J_k(\xib)$ is the Lagrange polynomial satisfying $J_k(\xib)\in \Gamma_I$ and $J_k(\xib_j)=\delta_{kj}$. Denoting $\hat{\ub}(\xb,\xib):=\sum_{k=1}^Q \ub(\xb,\xib_k)J_k(\xib)$, the collocation procedure to solve the stochastic nonlocal equation is $R(\hat{\ub}(\xb,\xib))|_{\xib_k}=0$, $\forall k=1,\cdots,Q$, where $R$ is the residual of \eqref{eqn:nonlocal_random}. With the property of Lagrange interpolation, we obtain
\begin{equation}\label{eqn:lag}
\left\{\begin{array}{ll}
\mcL_{H\delta}\ub(\xb,\xib_k) = \fb(\xb),&\quad \text{ in }\omg\\
\theta(\xb,\xib_k)=\int_{B_\delta (\xb)} K(\left|\yb-\xb\right|) (\yb-\xb)^T \left(\ub(\yb,\xib_k) - \ub(\xb,\xib_k) \right)d\yb,&\quad \text{ in }\omg\cup\omgb\\
\ub(\xb,\xib_k)=\ub_D(\xb,\xib_k), &\quad \text{ in }\omgbb
\end{array}\right.
\end{equation}
for $k=1,\cdots,Q$. Note that \eqref{eqn:lag} is equivalent to solving $Q$ deterministic nonlocal peridynamics problems, where the deterministic meshfree solver discussed in Section \ref{sec:meshfree} can be readily applied. Therefore, the PCM approach can be implemented in an embarrassingly parallel way and the total computational cost is the product of the number of collocation points and the cost of solving a deterministic problem.

To choose the set of prescribed collocation nodes $\Theta_N$, in this work we consider two different strategies: the tensor products of 1D collocation point sets and a sparse grid strategy for high dimensionality. In the tensor product strategy, one first construct a 1D interpolation for each dimension in the random space. For the $i$-th dimension, we take $\varpi_{(i)}$ numbers of nodal points $\Theta^{\varpi_{(i)}}_1=\{\xi_{1}^i,\cdots,\xi_{\varpi_{(i)}}^i\}\subset[-1,1]$, a 1D interpolation for a smooth function $v$ on the $i$-th dimension then writes:
\begin{equation}
  \mathcal{U}^{\varpi_{(i)}}[v](\xi_{(i)})=\sum_{k=1}^{\varpi_{(i)}}v(\xi^i_k)J^i_k(\xi_{(i)})
\end{equation}
where $J^i_k(\xi_{(i)})$ is the 1D Lagrange polynomial. Then for the case with high dimensionality in parametric space $v:\real^N\rightarrow\real$, the tensor product formula is:
\begin{equation}\label{eqn:tensor}
    \mathcal{J}[v]=\left(\mathcal{U}^{\varpi_{(1)}}\otimes\cdots\otimes\mathcal{U}^{\varpi_{(N)}}\right)[v]=\sum_{k_1=1}^{\varpi_{(1)}}\cdots\sum_{k_N=1}^{\varpi_{(N)}} v\left(\xi^1_{k_1},\cdots,\xi^N_{k_N}\right)\left(J^1_{k_1}\otimes\cdots\otimes J^N_{k_N}\right).
\end{equation}
Notice here \eqref{eqn:tensor} requires $Q=\Pi_{i=1}^{N}\varpi_{(i)}$ numbers of collocation points in total, which grows exponentially as $N$ increases and makes the simulation non-feasible (see, e.g., \cite{lin2009efficient}).
Therefore, the tensor product strategy may be employed for problems with a small number of random dimension.
For problems with a relatively large random dimension, we employ the sparse grids strategy. In particular, we employ the sparse grids constructed by the Smolyak algorithm \cite{smolyak1963quadrature}, which is a linear combination of tensor product formulas:
\begin{equation}\label{eqn:sparsegrid}
  \mathcal{J}[v]=\sum_{\zeta-N+1\leq|\bm{\varpi}|\leq \zeta}(-1)^{\zeta-\verti{\bm{\varpi}}_{l_1}}\binom{N-1}{\zeta-\verti{\bm{\varpi}}_{l_1}}\left(\mathcal{U}^{\varpi_{(1)}}\otimes\cdots\otimes\mathcal{U}^{\varpi_{(N)}}\right).
\end{equation}
Here $\zeta$ is the sparseness parameter, $\bm{\varpi}=(\varpi_{(1)},\cdots,\varpi_{(N)})\in\mathbb{N}^N$, $\verti{\bm{\varpi}}_{l_1}=\sum_{i=1}^N \varpi_{(i)}$, and $\varpi_{(i)}$ represents the number of collocation points in random dimension $i$. To compute \eqref{eqn:sparsegrid}, only evaluations on the sparse grids are needed:
\begin{equation}\label{eqn:sparsesample}
\Theta_N=\underset{\zeta-N+1\leq|\bm{\varpi}|_{l_1}\leq \zeta}{\bigcup} \left(\Theta^{\varpi_{(1)}}_1\times\cdots\times\Theta^{\varpi_{(N)}}_1\right).
\end{equation}
As shown in \cite{novak1996high,novak1999simple}, \eqref{eqn:sparsegrid} is exact for $p(\xib)\in\mathbb{P}_{\zeta-N}(\real^N)$ (all polynomials of degree less than $\zeta-N$) and the total number of nodes $Q\sim \frac{{2N}^{\zeta-N}}{(\zeta-N)!}$. Therefore, we may see that the sparse grids formulation typically requires a much smaller number of collocation points $Q$ than the full tensor product set and we will refer $\eta=\zeta-N$ as the ``level'' of the Smolyak formulation. As suggested in \cite{lin2009efficient}, generally the tensor product strategy is employed when the dimension of parametric space $N\leq4$, and the Smolyak sparse grid is preferred when $N>4$.

With a proper choice of $\Theta_N$, the statistical moments of each component of the random solution can then be evaluated with the numerical solution of \eqref{eqn:lag} on all probabilisitic collocation points $\xib_k\in\Theta_N$.  
To numerically compute the mean and the standard deviation of any function $q(\xb,\xib)$ of interest, we employ the quadrature rule approximation by choosing the set $\Theta_N$ as quadrature point set:
\begin{align}
&\mathbb{E}[q](\xb)\approx \sum_{k=1}^Q q(\xb,\xib_k)\mu_k,\label{eqn:E}\\
&\sigma[q](\xb)\approx\sqrt{\sum_{k=1}^Q (q(\xb,\xib_k))^2\mu_k-\left[\sum_{l=1}^Q q(\xb,\xib_l)\mu_l\right]^2},\label{eqn:std}
\end{align}
where $\{\mu_k\}_{k=1}^Q$ is the set of corresponding quadrature weights.



We now investigate the approximation error of PCM in the parametric space.
First, by our assumptions, $\lambda(\xb,\yb,\xib)$ and $\mu(\xb,\yb,\xib)$ are continuous in $\xib\in \Gamma$. Therefore, using similar arguments presented in the last part of Theorem \ref{thm:compatibility}, one can easily see that the map $\ub_\delta(\cdot,\xib): \Gamma \mapsto S_{H \delta}(\omg)$ is continuous, i.e., $\ub_\delta\in C(\Gamma; S_{H \delta}(\omg))$.  
Next, we follow the error analysis in \cite{nobile2008sparse}, which depends on higher regularity of the solution with respect to the parameter $\xib\in\Gamma$. We make the following regularity assumption for the rest of this subsection. 

\begin{assumption}[regularity]
\label{assu:regularity}
For each $\delta$,  we assume that the map $\ub_\delta(\cdot,\xib):\Gamma \mapsto S_{H \delta}(\omg)$   admits an analytic extension to the region
$
\mcA(\Gamma, \tau):=\{ \hat\xib\in \C^N : \dist(\hat\xib, \Gamma)\leq \tau \}
$. Moreover,
\[
\max_{\hat\xib\in\mcA(\Gamma, \tau) }\| \ub_\delta(\cdot,\hat\xib)\|_{S_{H \delta}(\omg; \C^d)} \leq C
\]
for some $C>0$. Note that the space $S_{H \delta}(\omg; \C^d)$ is defined by 
\[
S_{H \delta}(\omg; \C^d):= \{\ub\in L^2(\omg;\C^d):  |\ub|_{S_{H \delta}(\omg; \C^d)}^2:= \iint_{(\omg\cup \omgbb)^2}  \dfrac{K(\left|\yb-\xb\right|)}{\left|\yb-\xb\right|^2}\left|\left(\yb-\xb\right)\cdot\left(\ub(\yb) - \ub(\xb) \right)\right|^2 d\bm y d\bm x< \infty \}
\]
where $|\vb|^2 $ is understood as $\overline{\vb}{\vb}$ for $\vb:\omg \to \C^d$. 
\end{assumption}

In \cite[Theorems 3.10-3.11]{nobile2008sparse}, error analysis of the Smolyak sparse grids is presented for the classical linear elliptic PDEs, which is based on a fundamental result on the polynomial approximation of analytic functions. Here we present a similar result of \cite[Lemma 3.2]{nobile2008sparse} (see also \cite[Lemma 4.4]{babuvska2007stochastic}) which is the key lemma for the convergence theorem.  
\begin{lemma}
\label{lem:poly_approx_err}
Let $\Gamma^1=[-1,1]$ and $\mathbb{P}_p$ denote the polynomial space of degree $p$. Given a function $\vb(\xb, t)\in C(\Gamma^1;S_{H \delta}(\omg) )$ which admits an analytic extension to the region $\mcA(\Gamma^1,\tau)= \{ z\in \C : \dist(z, \Gamma^1)\leq \tau \}$ for some $\tau>0$, then 
\[
\min_{\wb\in\mcP_p \otimes S_{H \delta}(\omg; \C^d)} \|\vb -\wb \|_{C(\Gamma^1;S_{H \delta}(\omg))}\leq \frac{2}{\varrho - 1} e^{-p \log(\varrho)} \max_{z\in\mcA(\Gamma^1, \tau) }\| \vb(\cdot,z)\|_{S_{H \delta}(\omg; \C^d)} 
\]
where $\varrho = 2\tau + \sqrt{1+ 4 \tau^2}$.
\end{lemma}
\begin{proof}
The inequality is shown by taking $\wb$ to be the truncated Chebyshev expansion of $\vb$ up to degree $p$ which follows the proof of \cite[Lemma 4.4]{babuvska2007stochastic}. Since our functions are vector valued, we show the proof of the inequality for completeness. Let $\{ T_k(t)\}_{k=1}^\infty$ be the Chebyshev polynomials on $[-1,1]$, then the expansion of $\vb(\xb, t)=\vb(\xb,\cos(t))$ in $t$ is given by  
\[
\vb(\xb, t) = \frac{\bm a_0(\xb)}{2} + \sum_{k=1}^\infty \bm a_k(\xb) T_k(t)
\]
where $\bm a_k \in S_{H \delta}(\omg)$, $k=0,1, \cdots,$ are given by 
\[
\bm a_k(\xb) = \frac{1}{\pi} \int_{-\pi}^\pi \vb\big(\xb, \cos(s)\big) \cos(ks) ds. 
\]
The Chebyshev series has an analytic extension which converges in any open elliptic disc delimited by the ellipse $E_{\varrho}$ with foci $\pm 1$ and the sum of the half-axes $\varrho$ (see e.g. \cite{devore1993constructive}). Let $\wb =a_0(\xb)/2 + \sum_{k=1}^p \bm a_k(\xb) T_k(t)$, then
\[
\|\vb -\wb \|_{C(\Gamma^1;S_{H \delta}(\omg))} \leq \sum_{k=q+1}^\infty \| \bm a_k\|_{S_{H \delta}(\omg)}=\sum_{k=q+1}^\infty \| \bm a_k\|_{S_{H \delta}(\omg;\C^d)}. 
\]
Now for any $\hat\varrho$ with $1<\hat\varrho < \varrho$, following the arguments of \cite[Chapter 7, Theorem 8.1]{devore1993constructive}, one can rewrite $\bm a_k$ as 
\begin{equation}
\label{eqn:a_k_1}
\bm a_k(\xb) = \frac{1}{2\pi i} \int_{C_1} \vb\left(\xb, \frac{z+z^{-1}}{2}\right) z^{k-1} dz  + \frac{1}{2\pi i} \int_{C_2} \vb\left(\xb, \frac{z+z^{-1}}{2}\right) z^{-k-1} dz
\end{equation}
where $C_1:= \{ z\in \C: |z| =\hat\varrho^{-1}\}$ and $C_2:= \{ z\in \C: |z| =\hat\varrho\}$. 
Now we do change of variables with $z= \hat\varrho^{-1} e^{is}$ 
for the first integral in \eqref{eqn:a_k_1} and $z= \hat\varrho e^{is}$ for the second integral in \eqref{eqn:a_k_1}, we get 
\begin{equation}
  \label{eqn:a_k_2}
\bm a_k(\xb) = \frac{1}{2\pi } \int_{-\pi}^\pi \vb\left(\xb, \hat\varrho^{-1}\cos(s)\right) \hat\varrho^{-k} e^{iks}ds  + \frac{1}{2\pi} \int_{-\pi}^\pi \vb\left(\xb, \hat\varrho\cos(s)\right) \hat\varrho^{-k} e^{-iks}ds. 
\end{equation}
Using \eqref{eqn:a_k_2}, it is then easy to see that 
\[
\| \bm a_k\|_{S_{H \delta}(\omg;\C^d)} \leq 2 \hat\varrho^{-k} \max_{z\in \mcA(\Gamma^1,\tau)}\| \vb(\cdot, z)\|_{S_{H \delta}(\omg;\C^d)}.  
\]
So
\[\|\vb -\wb \|_{C(\Gamma^1;S_{H \delta}(\omg))} 
\leq \sum_{k=q+1}^\infty \| \bm a_k\|_{S_{H \delta}(\omg;\C^d)} \leq \frac{2}{\hat\varrho -1} \hat\varrho^{-p} \max_{z\in \mcA(\Gamma^1,\tau)}\| \vb(\cdot, z)\|_{S_{H \delta}(\omg;\C^d)}.
\]
Taking $\hat\varrho \to \varrho$, we get the desired result. 
\end{proof}

Once we have Lemma \ref{lem:poly_approx_err}, which is an analogue of \cite[Lemma 3.2]{nobile2008sparse}, we can conclude with the following convergence theorem. The proof is omitted since it follows the arguments in \cite[Theorems 3.10-3.11]{nobile2008sparse}.  

\begin{theorem}\label{thm:pcm_conv}
Assume that $\ub_\delta$ satisfies Assumption \ref{assu:regularity}.
Let $\ub_\delta^Q(\xb, \xib) = \sum_{k=1}^Q \ub_\delta(\xb,\xib_k)J_k(\xib)$. There exists $C_1>0$ and $\beta_1>0$ depending on $N$ and the analytic region $\mcA(\Gamma, \tau)$ such that  
 \begin{equation}
 \label{eqn:algebraic}
\max_{\xib\in\Gamma}\|\ub_\delta(\cdot,\xib) - \ub_\delta^Q(\cdot,\xib)\|_{S_{H\delta}(\omg)}  \leq C_1 Q^{-\beta_1}.
 \end{equation}
 Moreover, when $\eta > \frac{N}{\log(2)}$, there exists $C_2>0$, $C_3>0$ and $\beta_2>0$ depending on $N$ and the analytic region $\mcA(\Gamma, \tau)$, and $\beta_3>0$ depending only on $N$ such that 
\begin{equation}
 \label{eqn:subexponential}
\max_{\xib\in\Gamma}\|\ub_\delta(\cdot,\xib) - \ub_\delta^Q(\cdot,\xib)\|_{S_{H\delta}(\omg)}   \leq C_2 Q^{\beta_2} e^{- C_3 Q^{\beta_3}}.
 \end{equation}
\end{theorem}
\begin{remark}\label{rmk:tensor}
The convergence of the sparse grid approximation in the parameter space is presented in Theorem \ref{thm:pcm_conv}  as the number of $Q$ increases. If we instead use the tensor product formula \eqref{eqn:tensor} with 1D Chebyshev points for each dimension in the parameter space (then $p_{(i)}=\varpi_{(i)}-1$ for the $i$-th dimension, and the total number of samples $Q=\Pi_{i=1}^{N}\varpi_{(i)}$), then one can use the one dimensional result presented in Lemma \ref{lem:poly_approx_err} to get a convergence order. In particular, if we assume $\varpi= \varpi_{(i)}$ for $i=1, 2,\cdots, N$, then we have a convergence order $O( e^{-\varpi \log(\varrho)} ) =O( e^{-Q^{1/N} \log(\varrho)} ) $ where $\varrho$ depends on the analytic region $\mcA(\Gamma, \tau)$.
\end{remark}
We now present a result on the estimate of the difference between $\ub_\delta^Q$ and $\ub_0$. 
\begin{theorem}
\label{thm:pcm_conv_local}
Assume that $\ub_0$ satisfies Assumption \ref{assu:regularity} with $\delta=0$. Then there exists $C_1>0$ and $\beta_1>0$ depending on $N$ and the analytic region $\mcA(\Gamma, \tau)$ such that    
\begin{equation}
\max_{\xib\in\Gamma}\|\ub_0(\cdot,\xib) - \ub_\delta^Q(\cdot,\xib)\|_{S_{H\delta}(\omg)}\leq \Lambda(\eta, N) \max_{\xib\in\Gamma}\| \ub_0(\cdot,\xib) - \ub_\delta(\cdot,\xib)\|_{S_{H\delta}(\omg)} + C_1 Q^{-\beta_1}. 
\end{equation} 
Moreover, when $\eta > \frac{N}{\log(2)}$, there exists $C_2>0$, $C_3>0$ and $\beta_2>0$ depending on $N$ and the analytic region $\mcA(\Gamma, \tau)$, and $\beta_3>0$ depending only on $N$ such that \begin{equation}
\max_{\xib\in\Gamma}\|\ub_0(\cdot,\xib) - \ub_\delta^Q(\cdot,\xib)\|_{S_{H\delta}(\omg)}\leq \Lambda(\eta, N)\max_{\xib\in\Gamma} \| \ub_0(\cdot,\xib) - \ub_\delta(\cdot,\xib)\|_{S_{H\delta}(\omg)} +C_2 Q^{\beta_2} e^{- C_3 Q^{\beta_3}}. 
\end{equation} 
$\Lambda(\eta, N)$ is the Lebesgue constant associated with the sparse grid interpolation, satisfying
\begin{equation}
\label{eqn:Lebesgue}
\Lambda(\eta, N)\leq \sum_{\zeta-N+1\leq|\bm{\varpi}|_{l_1}\leq \zeta}\binom{N-1}{\zeta-\verti{\bm{\varpi}}_{l_1}}\prod_{j=1}^N \left(\frac{2}{\pi} \log(\varpi_{(j)}+1) +1\right).    
\end{equation}
\end{theorem}
\begin{proof}
Let $\ub_0 - \ub_\delta^Q = \ub_0 - \ub_0^Q +\ub_0^Q -\ub_\delta^Q $. The term $\max_{\xib\in\Gamma}\|\ub_0(\cdot,\xib) - \ub_0^Q(\cdot,\xib)\|_{S_{H\delta}(\omg)}$ can then be estimated by Theorem \ref{thm:pcm_conv}. 
Notice that  $\Lambda(\eta, N)$ be the Lebesgue constant associated with the sparse grid interpolation, i.e.,  
\[
\Lambda(\eta, N) := \sup_{v\in C(\Gamma)}\frac{\|\mathcal{J}[v]\|_{L^\infty}}{\| v\|_{L^\infty}},
\]
where $\mathcal{J}[v]$ is given by \eqref{eqn:sparsegrid}, then we have 
\[
\max_{\xib\in\Gamma}\|\ub_0^Q(\cdot,\xib) - \ub_\delta^Q(\cdot,\xib)\|_{S_{H\delta}(\omg)} \leq \Lambda(\eta, N) \max_{\xib\in\Gamma}\| \ub_0(\cdot,\xib) - \ub_\delta(\cdot,\xib)\|_{S_{H\delta}(\omg)},
\]
which leads to the desired results. 
\end{proof}

\subsection{Stochastic Peridynamics Formulation with Fracture}\label{sec:stochstic_LPS}

We now extend the optimization-based quadrature rule introduced in Section \ref{sec:meshfree} to the stochastic LPS model with fracture. 

For a given point $\xb_i$ and the horizon $\delta$, a bond is associated with each neighbor $\xb_j\in B_{\delta}(\xb_i)$, and the weight $\omega_{j,i}$ is associated with this bond. In the meshfree formulation, the fracture surface and the corresponding Neumann-type boundary $\partial\omg_N$ is represented by breaking bonds between $\xb_i$ and $\xb_j\in B_\delta(\xb_i)\backslash\omg_D$. For $\xb_j\in B_\delta(\xb_i)\cap\omg_D$ and when their bond stretch has not exceeded the critical bond stretch described in \eqref{eq:criteria}, we denote the bond between $\xb_i$ and $\xb_j$ as ``intact'' and the change of displacement on material point $\xb_j$ may have an impact on the displacement at $\xb_i$. On the other hand, when $\xb_j\notin \omg_D$ and/or when $s(\xb,\yb,\tau,\xib)>s_0(\xb,\yb,\xib)$ for some time $\tau<t$, we consider the bonds between $\xb_i$ and $\xb_j$ as ``broken''. To discretize the LPS formulation \eqref{eq:newform2}-\eqref{eq:continuousNonlocdilatation3_new_f}, the quadrature weights associated with intact bonds will be employed in the calculation of integrals inside $B_\delta(\xb_i)\cap\omg_D$ and the weights associated with broken bonds will be employed for integrals inside $B_\delta(\xb_i)\backslash\omg_D$. Particularly, we express the quadrature weights associated with intact bonds as $\tilde{\omega}_{j,i}$ and the quadrature weights associated with broken bonds as $\hat{\omega}_{j,i}$ through the scalar boolean state function $\gamma$. 
In particular, for each sample $\xib_k\in\Theta_N$, at the $n-$th step we set:
\begin{align}\label{eqn:gamma_disc}
     &\gamma^{n}_{j,i,k} = \begin{cases}
    1, \quad \text{if }\xb_j\in B_\delta(\xb_i)\cap\Omega_D \text{ and }s(\xb_i,\xb_i,t^l,\xib_k)\leq s_0(\xb_i,\xb_j,\xib_k),\;\forall l=1,\cdots,n,\\
    0, \quad \text{otherwise}, \\
    \end{cases}\\
    &{\tilde{\omega}^n_{j,i,k}}:={\omega}_{j,i}\gamma^n_{j,i,k}, \quad {\hat{\omega}^n_{j,i,k}}:={\omega}_{j,i}(1-\gamma^n_{j,i,k}). 
\end{align}
Notice that the new crack forms new free surfaces, which will be included in $\partial\omg_N$. Therefore, the computational domain $\omg$ will be updated with the evolution of cracks, we therefore denote the updated domain $\omg$ after the $n-$th step as $\omg^n$ and all subdomains such as $\omg_D$ will also be denoted with a similar fashion. Numerical quadrature of a given function $a(\xb)$ over $B_\delta(\xb_i)\cap\omg^n_D$ and $B_\delta(\xb_i)\backslash\omg^n_D$ may thus be calculated via
$$\int_{B_\delta(\xb_i)\cap\omg^n_D}a(\yb)d\yb\approx\sum_{\xb_j \in \chi_h\cap B_\delta(\xb_i)} \tilde{\omega}^n_{j,i,k} a(\xb_j),\qquad \int_{B_\delta(\xb_i)\backslash\omg^n_D}a(\yb)d\yb\approx\sum_{\xb_j \in \chi_h\cap B_\delta(\xb_i)} \hat{\omega}^n_{j,i,k} a(\xb_j).$$
This process is consistent with how damage is typically induced in bond-based peridynamics, such as the prototype microelastic brittle model \cite{silling_2005_2}. 

Applying the above formulation in \eqref{eq:newform2}-\eqref{eq:continuousNonlocdilatation3_new_f}, at the $n-$th quasi-static step, we aim to solve for the displacement $u^n_{i,k}\approx \ub(\xb_i,t^n,\xib_k)$ and nonlocal dilitation $\theta^n_{i,k}\approx \theta(\xb_i,t^n,\xib_k)$ through the following meshfree scheme:
{\begin{align}
   \nonumber (\mathcal{L}_{F\delta}^h \ub)^n_{i,k}=&\sum_{\xb_j \in \chi_h\cap B_\delta(\xb_i)}K_{ij}\left[\left(-\left(\lambda_{ij,k} - \mu_{ij,k}\right)  \left(\xb_j-\xb_i\right)\left(\theta^n_{i,k} + \theta^n_{j,k} \right)\right.\right.\\
 \nonumber &\left.\left.-8\mu_{ij,k} \frac{\left(\xb_j-\xb_i\right)\otimes\left(\xb_j-\xb_i\right)}{\left|\xb_j-\xb_i\right|^2} \cdot \left(\ub^n_{j,k}- \ub^n_{i,k} \right)\right) \tilde{\omega}^{n-1}_{j,i,k}\right.\\
  \nonumber&+\left(-{2\left(\lambda_{ij,k} - \mu_{ij,k}\right)}     \left(\xb_j-\xb_i \right) -4(\lambda_{ij,k}+2\mu_{ij,k})\mathbf{n}^{n-1}_{i,k}
     \frac{[\left(\xb_j-\xb_i \right)\cdot \mathbf{n}^{n-1}_{i,k}][\left(\xb_j-\xb_i \right)\cdot \mathbf{p}^{n-1}_{i,k}]^2}{\left|\xb_j-\xb_i\right|^2}\right.\\
     &\left.\left.+4\lambda_{ij,k}\mathbf{n}^{n}_{i,k}  \frac{[\left(\xb_j-\xb_i \right)\cdot \mathbf{n}^{n}_{i,k}]^3}{\left|\xb_j-\xb_i\right|^2}\right)\theta^n_{i,k}\hat{\omega}^{n-1}_{j,i,k}\right]= \fb(\xb_i),\label{eq:discreteNonlocElasticity2}\\
\theta^n_{i,k} =&\sum_{\xb_j \in \chi_h\cap B_\delta(\xb_i)} K_{ij} \left(\xb_j-\xb_i\right) \cdot \mathbf{M}^n_{i,k}\cdot \left(\ub^n_{j,k} - \ub^n_{i,k} \right) \tilde{\omega}^{n-1}_{j,i,k},\label{eq:discreteNonlocDilitation2}
\end{align}
where $\lambda_{ij,k}:=\lambda(\xb_i,\xb_j,\xib_k)$, $\mu_{ij,k}:=\mu(\xb_i,\xb_j,\xib_k)$,
\begin{equation}
\mathbf{M}^n_{i,k}:=\left[\sum_{\xb_j \in \chi_h\cap B_\delta(\xb_i)} K_{ij}(\xb_j-\xb_i)\otimes(\xb_j-\xb_i)\tilde{\omega}^{n-1}_{j,i,k}\right]^{-1},
\end{equation}}
the normal vector $\mathbf{n}({\xb})$ on free surfaces is numerically approximated and updated as
\begin{equation}\label{eqn:approx_n}
\nb^n_{i,k} = -\dfrac{\underset{\xb_j \in \chi_h\cap B_\delta(\xb_i)}{\sum}(\xb_j-\xb_i)\tilde{\omega}^{n-1}_{j,i,k} }{\vertii{\underset{\xb_j \in \chi_h\cap B_\delta(\xb_i)}{\sum}(\xb_j-\xb_i)\tilde{\omega}^{n-1}_{j,i,k}}},    
\end{equation}
and the tangential vector $\mathbf{p}^n_{i,k}$ is calculated as the orthogonal direction to $\nb^n_{i,k}$. The correction tensor should be invertible to ensure that the correction dilitation can be computed. This holds as long as the bonds in the horizon are non-colinear. For fracture case resulting in bond break, leaving an isolated particle, the matrix inverse may be replaced with the pseudo-inverse to improve the robustness.
To postprocess fracture evolution and identify cracks, the damage field $\phi_{i,k}^n\approx\phi(\xb_i,t^n,\xib_k)$ can then be defined as
\begin{equation}
{\phi_{i,k}^n}=\dfrac{\underset{\xb_j \in \chi_h\cap B_\delta(\xb_i)\setminus \xb_i}{\sum}(1-\gamma^n_{j,i,k})}{\underset{\xb_j \in \chi_h\cap B_\delta(\xb_i)\setminus\xb_i}{\sum}1},
\end{equation}
which indicates the weakening of material through the percentage of broken bonds in the neighborhood of $\xb_i$.

\section{Numerical Verification of Convergences}\label{sec:test}

In this section, we will investigate the asymptotic compatibility of the proposed method by testing the convergence of the numerical solution to the local limit. Three test problems are considered: a material deformation problem featuring smooth local limit for its displacement, a composite material deformation problem featuring discontinuous material properties, and an interfacial crack problem with in-plane extension of two dissimilar materials. In each test we study the $L^2$ errors for the mean and standard deviation of the solution. Let $\ub_{\delta}^{h,Q}$ represent the numerical solution with spatial grid size $h$ in meshfree methods and $Q$ samples in PCM, $\ub_0$ stands for the analytical local limit. We investigate the convergence of numerical solutions to the local limit as $Q$ increases and $\delta,h\rightarrow 0$ simultaneously with fixed ratio under the $\delta$-convergence limit. In particular we calculate the expectation $\mathbb{E}$ and standard derivation $\sigma$ 
\begin{equation}
\| {\mathbb{E}}(\ub_{\delta}^{h,Q})- \mathbb{E}(\ub_0) \| _{L_2(\omg)}, \quad \text{ and }\quad\|{\sigma}(\ub_{\delta}^{h,Q})- \sigma(\ub_0) \|_{L_2(\omg)}.
\end{equation}

In the stochastic problem, the Young's modulus $E(\xb,\xib)$ is set as a random field to represent the uncertainty in material microstructure, while Poisson ratio $\nu$ is taken as a constant in the whole domain.
Moreover, we assume that the material model satisfies the plane strain assumption: 
$$ \lambda(\xb,\xib)=E(\xb,\xib)\nu/((1+\nu)(1-2\nu)),\; \mu(\xb,\xib)=E(\xb,\xib)/(2(1+\nu)).$$ 
Following the conventions in \cite{yu2021asymptotically}, we adopt the nonlocal Lam{\'{e}} moduli as the harmonic mean of the local ones. Similarly, for problems with fracture, the local fracture energy $G(\xb,\xib)$ is also a random field, with the nonlocal fracture energy $G(\xb,\yb,\xib)$ defined via the arithmetic mean of the local ones. For all the tests in this section, the dimension $N$ of the parametric spaces is less than $4$. Therefore, in PCM the tensor product strategy is employed to generate the collocation point set $\Theta_N$. Moreover, in all numerical examples, we adopt the following popular scaled kernel for $K$:
\begin{equation}\label{eqn:K}
{K(r)}=\left\{\begin{array}{cl}
\dfrac{3}{\pi\delta^3r},\; &\text{ for }r\leq \delta; \\
0,\;& \text{ for }r> \delta. \\
\end{array}
\right.
\end{equation}  

\subsection{Test 1: a LPS problem with smooth local limit}

\begin{figure}[h!]
\centering
\subfigure[Convergence with $\delta,h\rightarrow 0$ in the physical space.]{\includegraphics[width=.45\columnwidth]{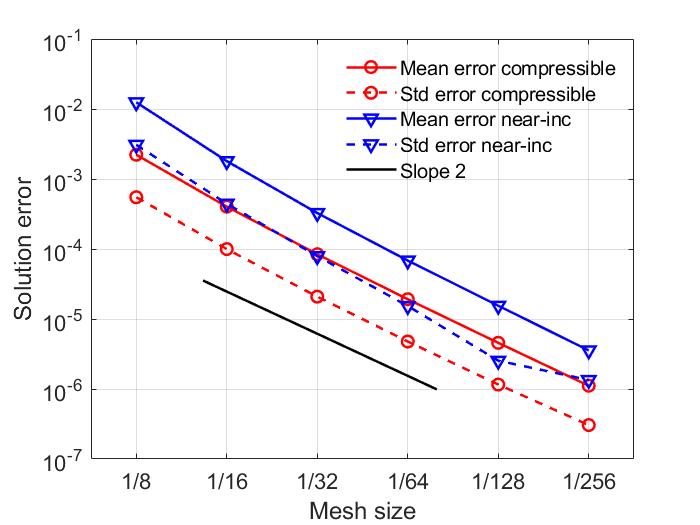}}\\
\subfigure[Convergence with sample numbers in the log scale.]{\includegraphics[width=.45\columnwidth]{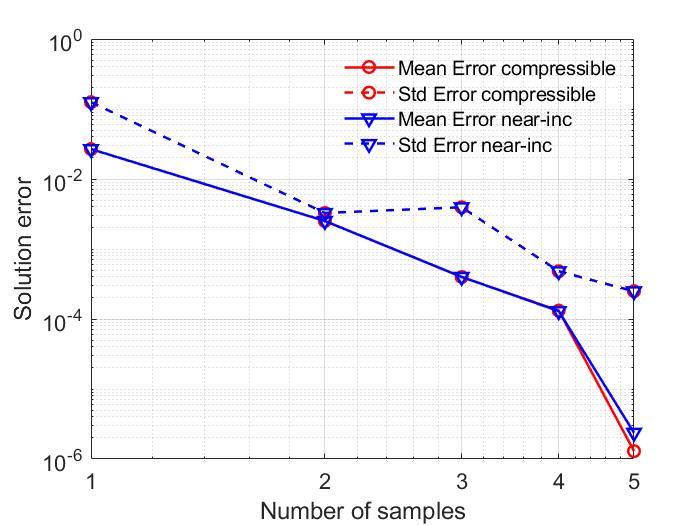}}
\subfigure[Convergence with sample numbers in the linear scale.]{\includegraphics[width=.45\columnwidth]{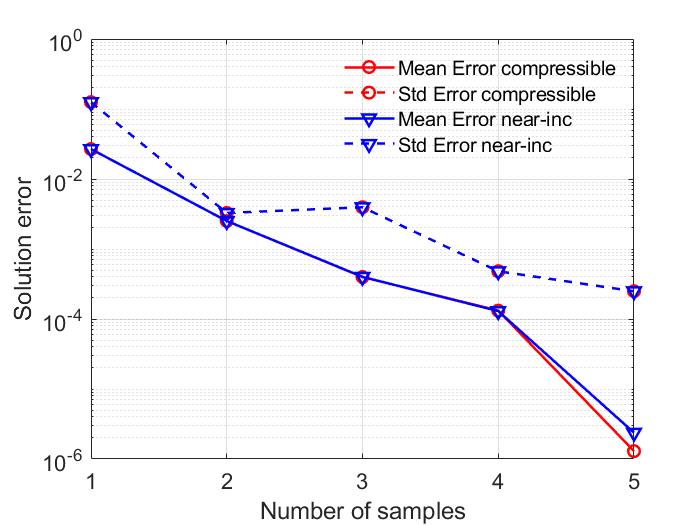}}
\caption{Convergence study of Test 1: a LPS problem with smooth local limit on 2D physical domain and 1D parametric space. Both compressible ($\nu=0.3$, as denoted by ``compressible'' cases) and nearly incompressible ($\nu=0.495$, as denoted by ``near-inc'' case) are investigated. Results in (a) are generated with $15$ samples. The data points in (b) and (c) are correspond to $1,2,...,5$ samples, respectively.}
\label{fig:test1}
\end{figure}

We first demonstrate the convergence rates on a Dirichlet-type LPS problem without fracture. In particular, we consider a case with 2D physical domain $\Omega=[-0.5,0.5]\times[-0.5,0.5]$ depending on a random variable $\xi$ following a Gaussian distribution $\xi\sim \mathcal{N}(0,0.1^2)$. The analytical local solution of displacement is given by
$$\ub_0(\xb,\xi)=\ub_0(x,y,\xi)=\left[\sin(x)\sin(y)/(2+\sin(5\xi)),-\cos(x)\cos(y)/(2+\sin(5\xi))\right], $$
with Young's modulus $$E(\xb,\xi)=E(x,y,\xi)=(2+\sin(x)\sin(y))(2+\sin(5\xi)),$$
and fixed loading
\begin{equation*}
\fb(\xb) =\fb(x,y)= \left[\begin{array}{c}
(C_1+C_2)(-4\sin(x)\sin(y)+2\cos(2x)\sin^2(y))+C_2(-4\sin(x)\sin(y)+2\cos(2y)\sin^2(x))\\
(C_1+2C_2)(4\cos(x)\cos(y)+\sin(2x)\sin(2y))
\end{array}\right]^T,
\end{equation*}
where $C_1:=\nu/((1+\nu)(1-2\nu)), C_2:=1/(2(1+\nu))$.
In this problem we consider full Dirichlet-type boundary condition on $\partial\omg$, and Dirichlet-type boundary conditions are applied on $\omgbb$ as $\ub_D(x,y,\xi)=\ub_0(x,y,\xi)$. 
Two values of Poisson ratio, $\nu=0.3$ and $0.495$, are investigated which correspond to compressible (as denoted by ``compressible'') and nearly-incompressible (as denoted by ``near-inc'') materials, respectively. Here we notice that when $\nu=0.3$, Assumption \ref{asp} is satisfied and we therefore have the $O(\delta^2)$ convergence to the local limit guaranteed by Theorem \ref{thm:compatibility_with_rate}. However, when the material is nearly-incompressible, Assumption \ref{asp} is not satisfied.

Numerical results are provided in Figure \ref{fig:test1}. With fixed ratio $\delta/h=3.0$ and $Q=15$ samples, in Figure \ref{fig:test1}(a) we show the error of numerical solution with respect to the analytical local limit for grid sizes $h=\{1/8, 1/16, 1/32, 1/64, 1/128, 1/256\}$. The optimal second-order convergence $O(\delta^2)$ is observed, which is consistent with Corollary \ref{thm:compatible} and the results in \cite{yu2021asymptotically}. In Figures \ref{fig:test1}(b) and \ref{fig:test1}(c) we fix $h=1/256$ and $\delta=3.0h$, and show the convergence of solution error with increasing number of samples $Q\in\{1,\dots,5\}$ in the parametric space. {In Figure \ref{fig:test1}(b), the horizontal axis is taken as $\varpi$ (notice that we have $\varpi=Q$, the number of samples, in this 1D case) in the logarithm scale to investigate if the solution error has algebraic convergence or not, while in Figure \ref{fig:test1}(c), the horizontal axis is taken as the polynomial order $\varpi$ in the linear scale to investigate the exponential convergence.} Almost exponential convergence is observed empirically, verifying the analysis of Remark \ref{rmk:tensor}. Similar convergence rates are observed in the compressible and nearly incompressible cases, which indicates that the conditions in Assumption \ref{asp} is a sufficient condition for the compatibility property but not a necessity.

\subsection{Test 2: composite material with discontinuous material properties}

\begin{figure}[h!]
\centering
\subfigure[Convergence with $\delta,h\rightarrow 0$ in the physical space.]{\includegraphics[width=.45\columnwidth]{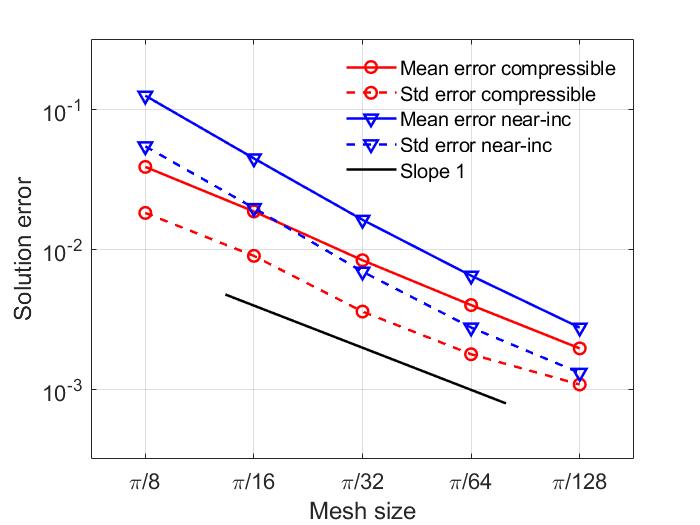}}\\
\subfigure[Convergence with sample numbers in the log scale.]{\includegraphics[width=.45\columnwidth]{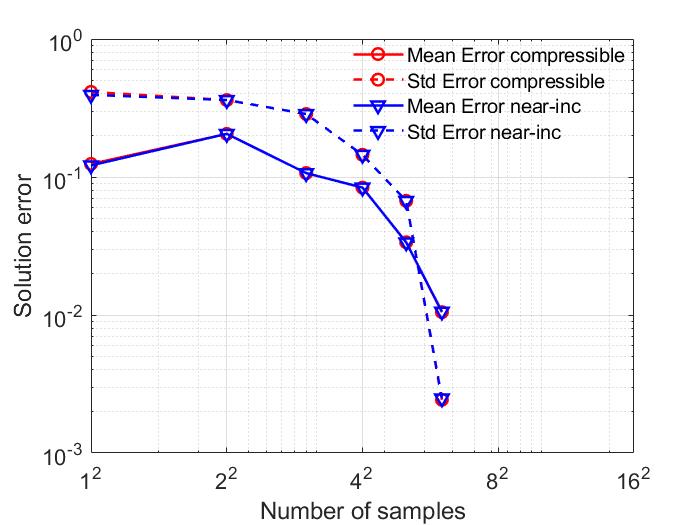}}
\subfigure[Convergence with sample numbers in the linear scale.]{\includegraphics[width=.45\columnwidth]{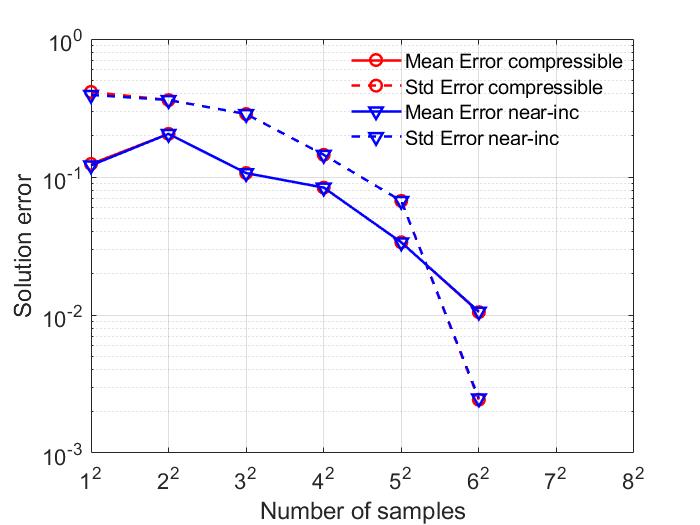}}
\caption{Convergence study of Test 2: composite material with discontinuous material properties on 2D physical domain and 2D parametric space. Both compressible ($\nu=0.3$, as denoted by ``compressible'' cases) and nearly incompressible ($\nu=0.495$, as denoted by ``near-inc'' cases) are investigated. Results in (a) are generated with $15^2=225$ samples. The data points in (b) and (c) are correspond to $1^2,2^2,...,6^2$ samples, respectively. }
\label{fig:test2}
\end{figure}

We now investigate composite materials with discontinuous material properties. A 2D physical domain $\Omega=[-\pi/2,\pi/2]\times[-\pi/2,\pi/2]$ and 2D parametric space $\xib=(\xi_{(1)},\xi_{(2)})$ are considered, where $\xi_{(1)},\xi_{(2)}$ are i.i.d. standard Gaussian random variables, i.e. $\xi_{(k)} \sim \mathcal{N}(0,1), k=1,2$. Denoting the left half of the physical domain as $\Omega_1:=[-\frac{\pi}{2},0]\times[-\frac{\pi}{2},\frac{\pi}{2}]$ and the right half as $\Omega_2:=[0,\frac{\pi}{2}]\times[-\frac{\pi}{2},\frac{\pi}{2}]$, the analytical local solution of displacement is given by
\begin{equation*}
\begin{aligned}
\ub_0(\xb,\xib)=&
\begin{cases}
\left[3x/(3+\sin(\xi_{(1)})+\sin(\xi_{(2)})),-x/(3+\sin(\xi_{(1)})+\sin(\xi_{(2)})\right], &\text{for }(x,y)\in \Omega_1\\
\left[1.5x/(3+\sin(\xi_{(1)})+\sin(\xi_{(2)})),-0.5x/(3+\sin(\xi_{(1)})+\sin(\xi_{(2)}))\right],  &\text{for }(x,y)\in \Omega_2\\
\end{cases}\\
\end{aligned}
\end{equation*}
with Young's modulus \begin{equation*}
\begin{aligned}
E(\xb,\xib)=&
\begin{cases}
3+\sin(\xi_{(1)})+\sin(\xi_{(2)}), &\text{ for }(x,y)\in \Omega_1\\
2(3+\sin(\xi_{(1)})+\sin(\xi_{(2)})), &\text{ for }(x,y)\in \Omega_2\\
\end{cases}
\end{aligned}
\end{equation*}
and zero loading forces $\fb$. In this example we also consider the LPS formulation with full Dirichlet-type boundary condition and without fracture. For $\xb\in\omgbb$, Dirichlet-type boundary conditions are applied as the analytical local solution. Similar as in Test 1, two values of Poisson ratio, $\nu=0.3$ and $0.495$, are investigated. Assumption \ref{asp} is satisfied when $\nu=0.3$, but not for $\nu=0.495$. Here we notice that with discontinuous material properties, the conditions in our compatibility Theorem \ref{thm:compatibility} is no longer satisfied. Therefore, with this example we aim to investigate the numerical stability and AC convergence rates that the theoretical analysis in Section \ref{sec:peri} does not cover. {On the other hand, with the smoothness of $\ub_0$ in the parametric space, it satisfies Assumption \ref{assu:regularity}. Therefore, as we increase $\varpi$ in PCM, an exponential convergence is expected from Remark \ref{rmk:tensor}.}

Numerical results are provided in Figure \ref{fig:test2}. With fixed ratio $\delta/h=3.0$ and {$Q=225$ samples}, in Figure \ref{fig:test2}(a) we show the error of numerical solution with respect to the analytical local limit for grid sizes $h=\{\pi/8, \pi/16, \pi/32, \pi/64, \pi/128\}$. First-order convergence $O(\delta)$ is observed, which is consistent with the numerical observations in \cite{yu2021asymptotically}. In Figures \ref{fig:test2}(b) and \ref{fig:test2}(c) we fix $h=\pi/256$ and $\delta=3.0h$, and show the convergence of solution error with increasing {number of samples $Q\in\{1^2,\cdots,6^2\}$} in the parametric space. {Similar as in test 1, in Figure \ref{fig:test2}(b), the horizontal axis is taken as $\varpi$ (notice that we have $Q=\varpi^2$ in this case, since the tensor product formula is employed in PCM) in the logarithm scale while in Figure \ref{fig:test2}(c), the horizontal axis is taken as $\varpi$ in the linear scale to investigate the exponential convergence.} An exponential convergence is observed empirically, {verified the analysis in Remark \ref{rmk:tensor}.}

\subsection{Test 3: material fracture on a bimaterial interface}

\begin{figure}[h!]
\centering
\subfigure[Problem setting, where crack lies on a bi-material interface subjected to remote loading. ]{\includegraphics[width=.35\columnwidth]{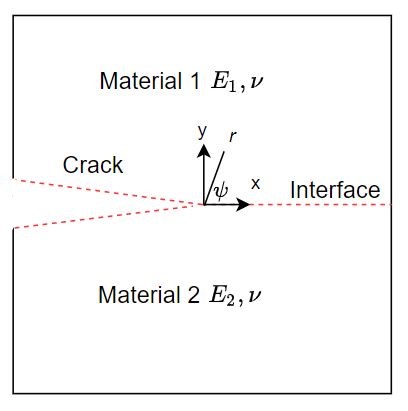}}
\subfigure[Analytical damage field.]{\includegraphics[width=.50\columnwidth]{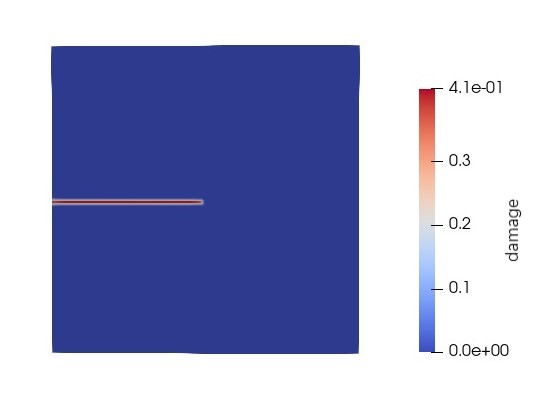}}
\subfigure[Analytical displacement on the $x$-direction.]{\includegraphics[width=.45\columnwidth]{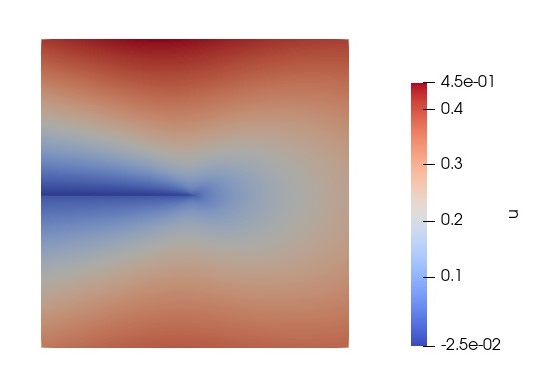}}
\subfigure[Analytical displacement on the $y$-direction.]{\includegraphics[width=.445\columnwidth]{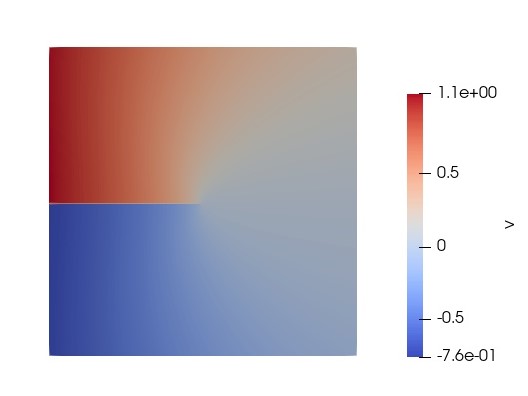}}
\caption{Problem setting and analytical solutions in Test 3: material fracture on a bimaterial interface.}
\label{fig:bicrack_u}
\end{figure}


In this example we proceed to consider the fracture problem. As shown in Figure \ref{fig:bicrack_u}, we consider the in-plane extension of two dissimilar materials with cracks along their interface. 
A physical domain $\Omega=[-\pi/2,\pi/2]\times[-\pi/2,\pi/2]$ is employed. The material property depends on a two i.i.d. random variables $\xi_{(1)}$ and $\xi_{(2)}$, where $\xi_{(1)}\sim \mathcal{N}(0,1)$ satisfies a Gaussian distribution and $\xi_{(2)}\sim \mathcal{U}[-1,1]$ satisfies a uniform distribution. The Young's modulus of the two materials, on the upper half plane and the lower half plane respectively, are denoted as $E_1(\xib)$ and $E_2(\xib)$. In particular, we take $E_1(\xib)=2+\sin(\xi_{(1)})$ and $E_2(\xib)=2+\sin(\xi_{(2)})$. Both compressible ($\nu=0.3$) and nearly incompressible ($\nu=0.495$) will be investigated. Again, Assumption \ref{asp} is satisfied when $\nu=0.3$, but not for $\nu=0.495$. For this problem the Cartesian component of the analytical local displacement field $\ub$ is given by \cite{wang2017xfem}:
\begin{equation}
\begin{split}
\ub_0(\xb,\xib)=\left[
\begin{array}{c} 
u(\xb,\xib) \\ 
v(\xb,\xib) 
\end{array}\right]^T=\sqrt{\dfrac{r(\xb)}{2\pi}}\left(\text{Re}(r(\xb)^{i\epsilon(\xib)})\left[\begin{array}{c} u^I(\psi(\xb),\xib) \\ v^I(\psi(\xb),\xib) \end{array}\right]+\text{Im}(r(\xb)^{i\epsilon(\xib)})\left[\begin{array}{c} u^{II}(\psi(\xb),\xib) \\ v^{II}(\psi(\xb),\xib) \end{array}\right]\right)^T
\end{split}
\label{eq:bicrack}
\end{equation}
where $(r(\xb),\psi(\xb))$ correspond to the local polar coordinate system of $\xb$ with origin at the crack tip, Re($\cdot$) and Im($\cdot$) denote the real and imaginary parts of a complex number, respectively. Notice that in this example we follow \cite{wang2017xfem} and take the complex stress intensity factor (SIF) as $1.0$. The bimaterial constant $\epsilon(\xib)$ depends on the material properties of both materials and leads to oscillation of near-tip displacements and stresses:
\begin{equation}
\epsilon(\xib)=\dfrac{1}{2\pi}\log{\dfrac{\mu_2(\xib)\kappa_1+\mu_1(\xib)}{\mu_1(\xib)\kappa_2+\mu_2(\xib)}},\quad \mu_m(\xib)=\dfrac{E_m(\xib)}{2(1+\nu)},\quad \kappa_m=3-4\nu,\text{ for }m=1,2.
\end{equation}
We set the material properties $\mu(\xb,\xib)=\mu_1(\xib),\kappa(\xb)=\kappa_1$ when $\xb$ is in the upper half-plan, and $\mu(\xb,\xib)=\mu_2(\xib),\kappa(\xb)=\kappa_2$ when $\xb$ is in the lower half-plan. ($u^I,v^I$) and ($u^{II},v^{II}$) are then functions of the angular $\psi(\xb)$ and $\xib$:
\begin{equation}
\begin{split}
u^I(\psi(\xb),\xib):=&-\dfrac{1}{2\mu(\xb,\xib)(1+4\epsilon(\xib)^2)\cosh(\pi\epsilon(\xib))}\Big\{ [e^{\epsilon(\xib)(\Pi(\xb)-\psi(\xb))}-\kappa(\xb) e^{-\epsilon(\xib)(\Pi(\xb)-\psi(\xb))}]\cos(\psi(\xb)/2)\\&-(1+4\epsilon(\xib)^2)e^{-\epsilon(\xib)(\Pi(\xb)-\psi(\xb))}\sin{\psi(\xb)}\sin(\psi(\xb)/2)\\
&+2\epsilon(\xib)[e^{\epsilon(\xib)(\Pi(\xb)-\psi(\xb))}+\kappa(\xb) e^{-\epsilon(\xib)(\Pi(\xb)-\psi(\xb))}]\sin(\psi(\xb)/2) \Big\},\\
v^I(\psi(\xb),\xib):=&\dfrac{1}{2\mu(\xb,\xib)(1+4\epsilon(\xib)^2)\cosh(\pi\epsilon(\xib))}\Big\{ [e^{\epsilon(\xib)(\Pi(\xb)-\psi(\xb))}+\kappa(\xb) e^{-\epsilon(\xib)(\Pi(\xb)-\psi(\xb))}]\sin{(\psi(\xb)/2)}\\&-(1+4\epsilon(\xib)^2)e^{-\epsilon(\xib)(\Pi(\xb)-\psi(\xb))}\sin{\psi(\xb)}\cos(\psi(\xb)/2)\\
&-2\epsilon(\xib)[e^{\epsilon(\xib)(\Pi(\xb)-\psi(\xb))}-\kappa(\xb) e^{-\epsilon(\xib)(\Pi(\xb)-\psi(\xb))}]\cos(\psi(\xb)/2) \Big\},\\
u^{II}(\psi(\xb),\xib):=&\dfrac{1}{2\mu(\xb,\xib)(1+4\epsilon(\xib)^2)\cosh(\pi\epsilon(\xib))}\Big\{ [e^{\epsilon(\xib)(\Pi(\xb)-\psi(\xb))}+\kappa(\xb) e^{-\epsilon(\xib)(\Pi(\xb)-\psi(\xb))}]\sin{({\psi(\xb)/2})}\\&+(1+4\epsilon(\xib)^2)e^{-\epsilon(\xib)(\Pi(\xb)-\psi(\xb))}\sin{\psi(\xb)}\cos(\psi(\xb)/2)\\
&-2\epsilon(\xib)[e^{\epsilon(\xib)(\Pi(\xb)-\psi(\xb))}-\kappa(\xb) e^{-\epsilon(\xib)(\Pi(\xb)-\psi(\xb))}]\cos(\psi(\xb)/2) \Big\},\\
v^{II}(\psi(\xb),\xib):=&\dfrac{1}{2\mu(\xb,\xib)(1+4\epsilon(\xib)^2)\cosh(\pi\epsilon(\xib))}\Big\{ [e^{\epsilon(\xib)(\Pi(\xb)-\psi(\xb))}-\kappa(\xb) e^{-\epsilon(\xib)(\Pi(\xb)-\psi(\xb))}]\cos(\psi(\xb)/2)\\&+(1+4\epsilon(\xib)^2)e^{-\epsilon(\xib)(\Pi(\xb)-\psi(\xb))}\sin{\psi(\xb)}\sin(\psi(\xb)/2)\\
&+2\epsilon(\xib)[e^{\epsilon(\xib)(\Pi(\xb)-\psi(\xb))}+\kappa(\xb) e^{-\epsilon(\xib)(\Pi(\xb)-\psi(\xb))}]\sin(\psi(\xb)/2) \Big\}.
\end{split}
\end{equation}
Here the value of $\Pi(\xb)$ also depends on the location of $\xb$: $\Pi(\xb)=\pi$ for $\xb$ on the upper half-plane, whereas $\Pi(\xb)=-\pi$ for the lower half-plane. In Figure \ref{fig:bicrack_u} we plot the analytical local solution for the damage field and the displacement fields for illustration. In particular, the crack is represented by breaking the bonds across the segment between $(-\frac{\pi}{2},0)$ and $(0,0)$. On the crack surface, free surface conditions are imposed, while full Dirichlet-type boundary conditions are applied on all four sides of the plate. 
Similar as in Test 2, in this example the Young's modulus $E(\xb,\xib)$ is (spatially) discontinuous across the interface, and therefore the conditions in our compatibility Theorem \ref{thm:compatibility} 
is no longer satisfied.

\begin{figure}[h!]
\centering
\subfigure[Convergence with $\delta,h\rightarrow 0$ in the physical space.]{\includegraphics[width=.45\columnwidth]{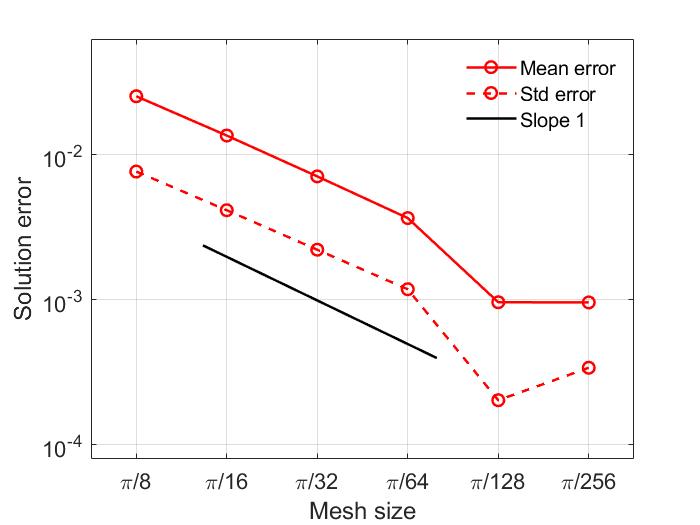}}\\
\subfigure[Convergence with sample numbers in the log scale.]{\includegraphics[width=.48\columnwidth]{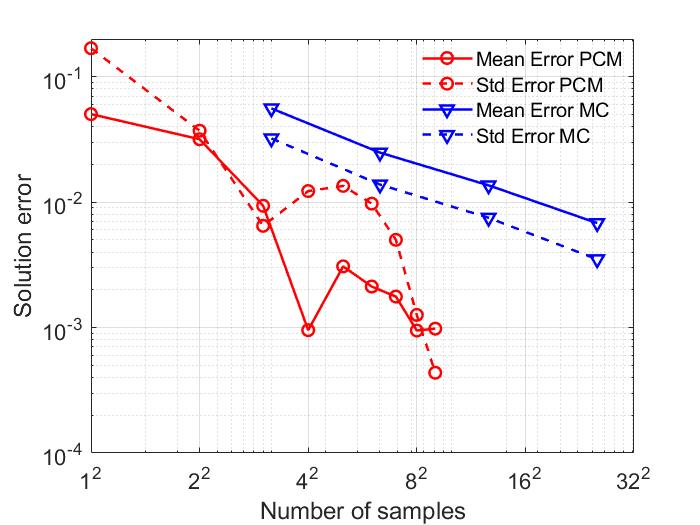}}
\subfigure[Convergence with sample numbers in the linear scale.]{\includegraphics[width=.48\columnwidth]{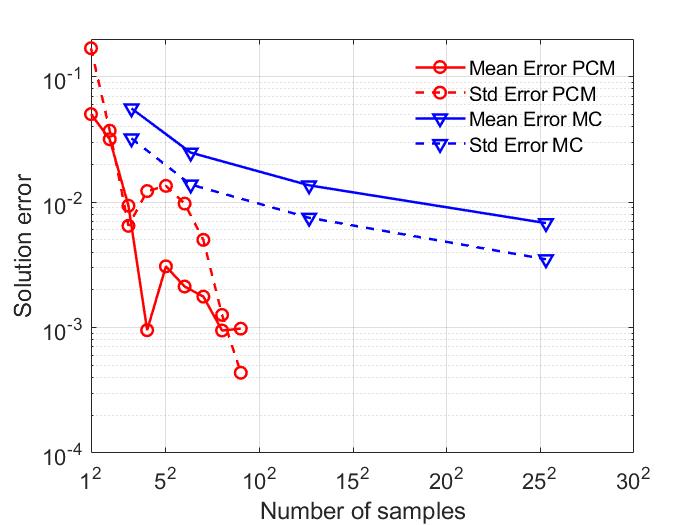}}
\caption{Convergence study of Test 3: material fracture on a bimaterial interface, for \textbf{compressible materials ($\nu$=0.3)}. Here we use ``PCM'' to denote the cases using our proposed probabilistic collocation method approach, and ``MC'' to denote the cases using the Monte Carlo method. Results in (a) are generated with $20^2=400$ samples. The data points in (b) and (c) are corresponding to $1^2,\cdots,9^2$ samples.
} 
\label{fig:bicrack_com}
\end{figure}

\begin{figure}[h!]
\centering
\subfigure[Convergence with $\delta,h\rightarrow 0$ in the physical space.]{\includegraphics[width=.45\columnwidth]{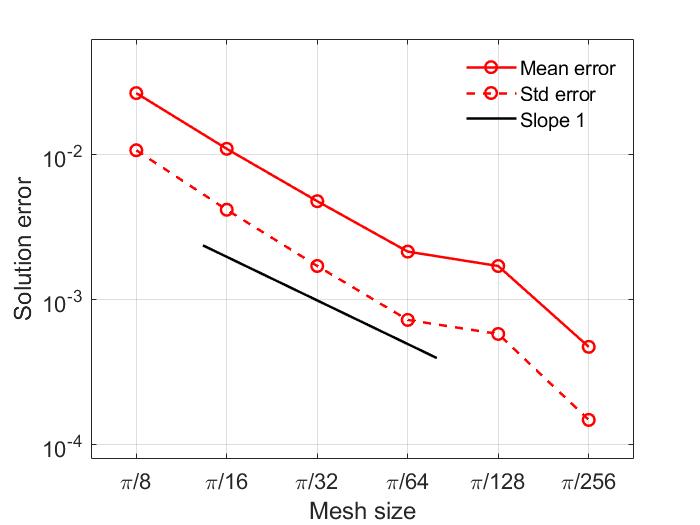}}\\
\subfigure[Convergence with sample numbers in the log scale.]{\includegraphics[width=.48\columnwidth]{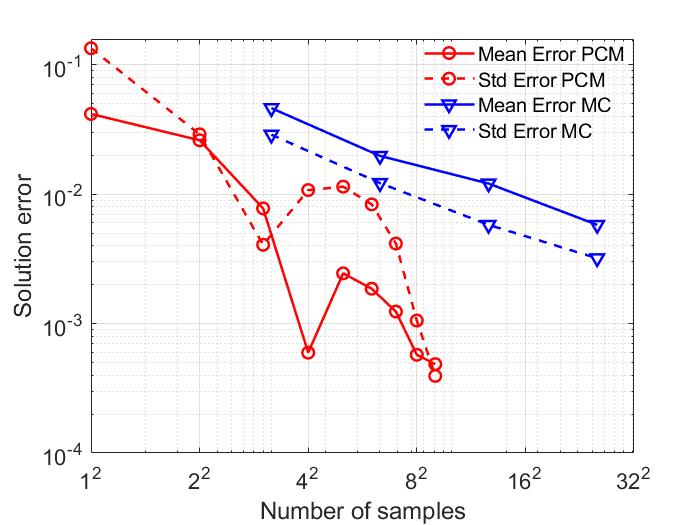}}
\subfigure[Convergence with sample numbers in the linear scale.]{\includegraphics[width=.48\columnwidth]{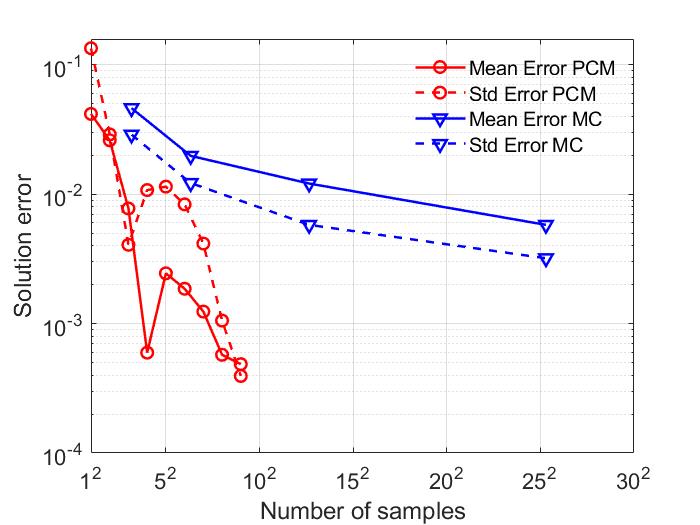}}
\caption{Convergence study of Test 3: material fracture on a bimaterial interface, for \textbf{nearly incompressible materials ($\nu$=0.495)}. Here we use ``PCM'' to denote the cases using our proposed probabilistic collocation method approach, and ``MC'' to denote the cases using the Monte Carlo method. Results in (a) are generated with {$20^2=400$} samples. The data points in (b) and (c) are corresponding to {$1^2,\cdots,9^2$} samples.}
\label{fig:bicrack_inc}
\end{figure}


Numerical results for compressible and nearly incompressible cases are provided in Figure \ref{fig:bicrack_com} and Figure \ref{fig:bicrack_inc}, respectively. With fixed ratio $\delta/h=3.0$ and $Q=400$ samples, in  Figure \ref{fig:bicrack_com}(a) and Figure \ref{fig:bicrack_inc}(a) we show the error of numerical solution with respect to the analytical local limit for grid sizes $h=\{\pi/8, \pi/16, \pi/32, \pi/64, \pi/128, \pi/256\}$. First-order convergence $O(\delta)$ is observed. In Figures \ref{fig:bicrack_com}(b), \ref{fig:bicrack_com}(c),
\ref{fig:bicrack_inc}(b) and 
\ref{fig:bicrack_inc}(c), using fixed grid size $h=\pi/256$ and $\delta=3.0h$, we demonstrate the convergence of solution errors with increasing {number of samples $Q=\{1^2,\cdots,9^2\}$} in the parametric space. {Similar as in tests 1 and 2, in Figures \ref{fig:bicrack_com}(b) and \ref{fig:bicrack_inc}(b), the error is plotted versus $\varpi$ in the logarithm scale while in Figures \ref{fig:bicrack_com}(c) and \ref{fig:bicrack_inc}(c) the horizontal axis is taken as $\varpi$ in the linear scale.} A roughly algebraic convergence rate is observed. We notice that the convergence curve seems more oscillatory comparing with the previous two tests, possibly due to the solution nonlinearity induced by the spatial discontinuity and the reduced regularity in the parametric space. {In fact, in \cite{foo2008multi,jakeman2013minimal,witteveen2013simplex}, a similar phenomenon of oscillatory convergence curve was observed, when the solution has discontinuity or reduced regularity in the parametric space (see, e.g., Figure 6 of \cite{Witteveen2013}).} To further demonstrate the sample efficiency of the proposed approach, we also plot the convergence of numerical solutions obtained from Monte Carlo (MC) simulations. The results indicate that to achieve a similar level of accuracy, our proposed approach requires a much smaller number of samples compared to MC. 

\section{Application: Brittle Fracture of Glass-Ceramics}\label{sec:val}

\begin{figure}[h!]
\centering
\subfigure{\includegraphics[width=.5\columnwidth]{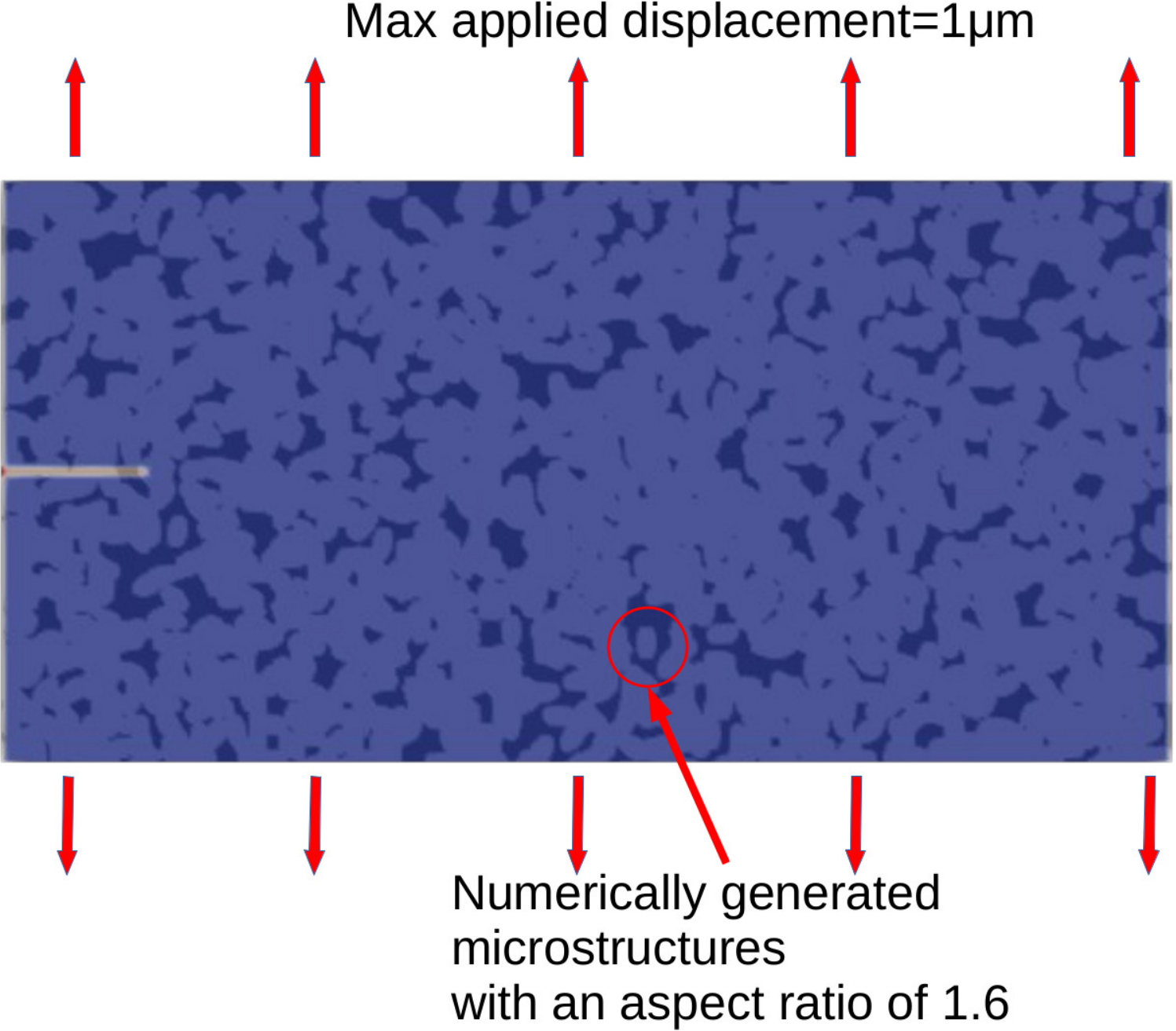}}
\caption{Problem setup of pre-cracked glass-ceramics experiment with randomly distributed material property fields, following \cite{serbena2015crystallization}. Here light blue represents the crystalline and dark blue represents the glassy matrix.}
\label{fig:glasssetup}
\end{figure}

\begin{table}[]
\center
\begin{tabular}{|c|cccc|}
\hline
&Young's modulus& Poisson ratio& Fracture energy& Fracture Toughness\\
\hline
Glass&$E_1=$80 $GPa$&0.25&$G_1=$6.59 $J/m^2$ & 0.75 $MPa\cdot m^{0.5}$\\
Crystal&$E_2=$133 $GPa$&0.25&$G_2=$86.35 $J/m^2$ & 3.5 $MPa\cdot m^{0.5}$\\
\hline
\end{tabular}
\caption{Material properties used in pre-cracked glass-ceramics experiment \cite{serbena2015crystallization}.}
\label{tab:glass}
\end{table}

Having illustrated the AC convergence convergence to the analytical local limits and verified the theoretical analysis in Sections \ref{sec:math}-\ref{sec:num}, we now consider {a problem of brittle fracture in a glass-ceramic material} as a prototypical exemplar, and provide validation against experiment results. The main objective of this section is to provide a proof-of-principle demonstration that the framework introduced thus far applies to realistic settings, however overall the provided preliminary validation provides good agreement. {A glass-ceramic material is the product of controlled crystallization of a specialized glass composition, which results in the creation of a microstructure composing of one of more crystalline phases within the residual amorphous glass}. {Glass-ceramics have received significant attention due to their enhanced strength and toughness compared to pure glass} \cite{prakash2022investigation, serbena2012internal,freiman1972effect, holand2019glass, fu2017nature}. {A wide range of flexural strength (100 to $\geq 500 MPa$) and fracture toughness (1.0 to 5.0 $MPa.m^{0.5}$) are reported in literature} \cite{fu2017nature}, {with the authors noting that the microstructure and phase assemblage play a vital role in determining strength and toughness}. Therefore, it is important to investigate the microstructure of these materials and their relation to damage metrics of interests to get fundamental insight \cite{serbena2015crystallization}. In particular, we employ the proposed approach to study the {fracture toughness of a model glass-ceramic material (lithium disilicate) as a function of crystal volume fraction} \cite{serbena2015crystallization}.

In this example, we consider a pre-notched {idealized microstructural realization} which is subject to displacement {boundary conditions} on its top and bottom {boundaries}, as demonstrated in Figure \ref{fig:glasssetup}. A plate of dimensions $800\mu m$ by $400\mu m$ is considered, with an initial crack of length $100\mu m$, and a gradually increasing displacement loading $U_D$ applied on the top and bottom of the sample. All other boundaries, including the new boundaries created by cracks, are treated as free surfaces. Each realization is composed {of randomly distributed crystals embedded in a glassy matrix}, with the mechanical properties of glass and crystalline phases listed in Table \ref{tab:glass}. In particular, we follow \cite{serbena2015crystallization,prakash2022investigation} and generate the center location  $(C_x,C_y)$ and rotation angle $C_\psi$ of each crystal as random variables satisfying $C_x\sim \mathcal{U}[0,800]$, $C_y\sim \mathcal{U}[0,400]$, and $C_\psi\sim \mathcal{U}[0,2\pi]$. All crystals are identical ellipses with semi-major and semi-minor axes being $12\mu m$ and $7.5\mu m$, respectively, with an aspect ratio of 1.6. This material was studied experimentally in \cite{serbena2015crystallization} for different crystallized volume fractions, $f$. Although the crack pattern varies drastically with different microstructure realizations, for each crystallized volume fraction $f$ the averaged fracture toughness presents a consistent pattern. In particular, a total of three samples were tested experimentally for each crystallized volume fraction and the average of these tests were reported in \cite{serbena2015crystallization}. It was observed that the averaged fracture toughness grows linearly with $f$. Therefore, in this example we aim to reproduce the experimental fracture toughness in \cite{serbena2015crystallization} rather than the individual crack pattern with numerical simulations, since the former is more reproducible and also provides a more direct measure of the material resistance.



\begin{figure}[h!]
\centering
\includegraphics[width=.95\columnwidth]{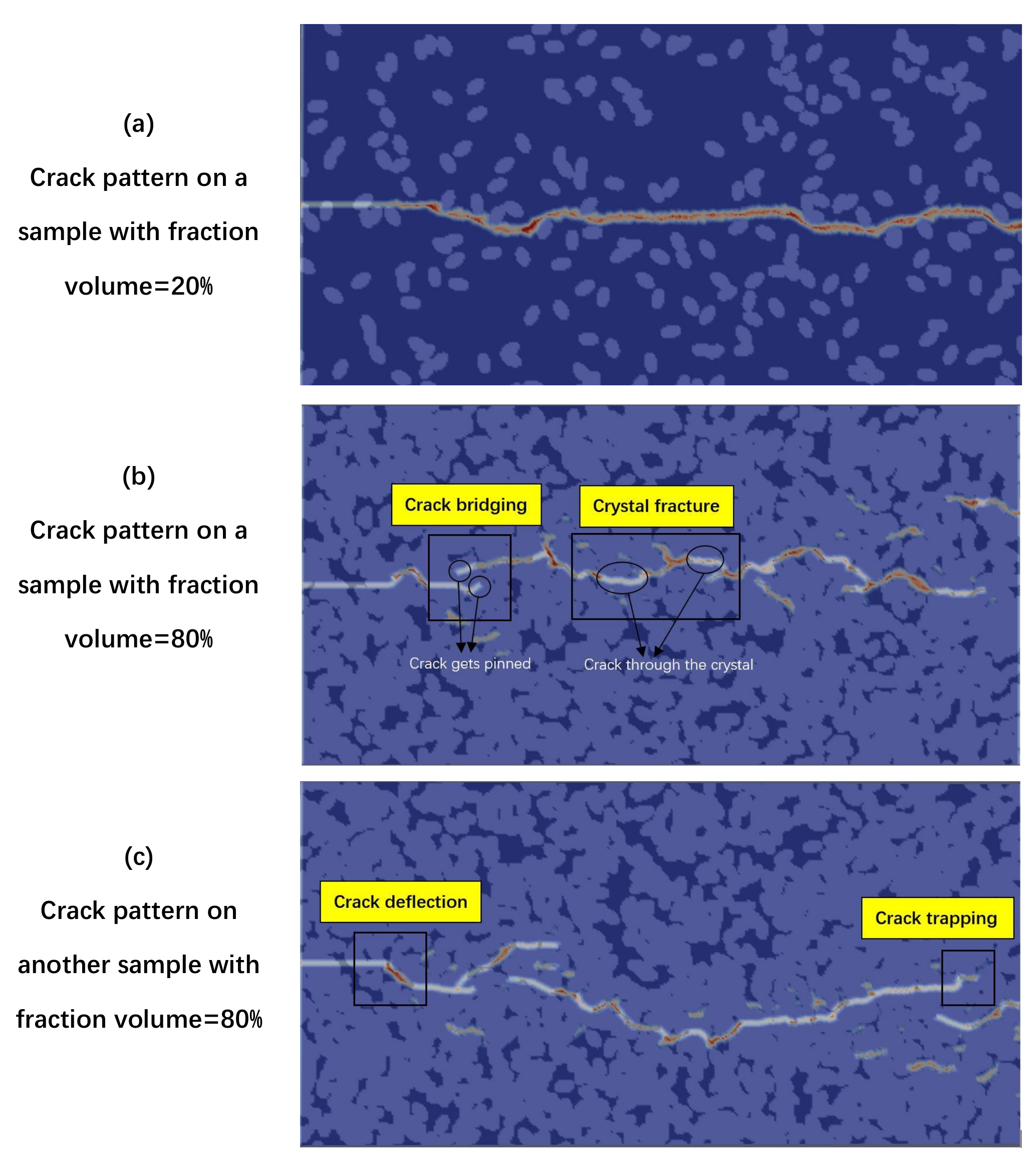}
\caption{Crack pattern (represented by the damage field $\phi$) of glass-ceramics on two sample microstructures. Here light blue represents the crystalline and dark blue represents the glassy matrix. (a) A sample with crystal volume fraction $f=20\%$. No crack bridging nor crystal fracture is observed. (b)(c) Two samples with crystal volume fraction $f=80\%$. The crack pattern is dominated by crystal fracture, and crack bridging, deflection and trapping are also observed.}
\label{fig:f2080}
\end{figure}

To numerically simulate the crack growth in this problem, we consider the plane strain model and employ the quasi-static LPS model setting as described in Section \ref{sec:stochstic_LPS}. In numerical experiments, we gradually increase $U_D$ from $0\mu m$ to $1\mu m$, and simulate the propagation of the crack starting from the pre-crack tip till it reaches the right boundary of the domain. At each quasi-static step, we increase $U_D$ by $4e-3\mu m$, perform subiterations until no new broken bonds are detected, then proceed to the next step. For spatial discretization, we employ uniform grids with grid size $h=2\mu m$, and the horizon size $\delta=3h=6\mu m$. Therefore, the whole computational domain $\omg\cup\omgbb$ has $M=87969$ grid points in total. Four different crystallized volume fraction values are considered: $20\%$, $40\%$, $60\%$ and $80\%$.

To demonstrate the performance of our deterministic LPS solver, in Figure \ref{fig:f2080} we show the crack pattern of two samples with volume fraction $20\%$ and $80\%$, respectively. In Figure \ref{fig:f2080}(a), one can observe that the {crack mostly propagates either inside the glassy matrix or along interfaces after crack deflection} and avoids entering the ceramic particles, on account of the fracture toughness of the ceramic phase being much higher. On the other hand, once we increase the crystallinity, as shown in Figure \ref{fig:f2080}(b)(c) where the crystals occupy $80\%$ of the volume, the crack pattern gets dominated by crystal fracture. {In certain cases, where a crack gets penetrates and gets trapped within a large agglomeration, it results in crack bridging wherein it is favourable for the crack to re-initiate in a nearby interface rather than fracturing the crystal agglomeration}.
Such patterns were also observed and reported in \cite{serbena2015crystallization}, where the authors considered crack deflection, trapping and bridging as the three main toughening mechanisms in glass-ceramics.


We now proceed to solve the stochastic LPS problem and provide a quantitative validation by comparing the numerical results on fracture toughness with the experimental measurements in \cite{serbena2015crystallization}. In this study, the material microstructure is treated as a random field, and the quantities of interest would be the averaged fracture toughness of different realizations for each volume fraction $f\in\{20\%,40\%,60\%,80\%\}$. For each realization, we use $R(\xb,\omega)$ to denote the microstructure, such that for each $\omega\in\Omega_p$, 
\begin{equation}
R(\xb,\omega)=\left\{\begin{array}{cc}
0     &\text{ if the material point $\xb$ is glass,}  \\
1     &\text{ if the material point $\xb$ is crystal.} 
\end{array}\right.
\end{equation}
We then notice that the random fields of Young's modulus $E(\xb,\omega)$ and fracture energy $G(\xb,\omega)$ can be represented as linear transformations of $R$:
$$E(\xb,\omega)=R(\xb,\omega)(E_2-E_1)+E_1,\quad G(\xb,\omega)=R(\xb,\omega)(G_2-G_1)+G_1,$$
where $E_1$, $E_2$ are the Young's modulus of glass and crystal, respectively, and $G_1$, $G_2$ are their fracture energy. The material responses and crack propagation in this sample can then be calculated using the LPS solver  \eqref{eq:discreteNonlocElasticity2}-\eqref{eq:discreteNonlocDilitation2}, and the fracture toughness is determined by the mechanisms through which cracks interact with constituents in microstructures \cite{li2013prediction}. Based on the final crack pattern, we first calculate the average energy release rate through
\begin{equation}
    G_{IC}=\dfrac{G_1L_1+G_2L_2+G_iL_i}{W},
\end{equation}
where $W$ is the total projected crack length along the $x$-direction and $G_i:=(G_1+G_2)/2$ denotes the the fracture energy for interface debonding. $L_1$, $L_2$ and $L_i$ are the crack length within the glass, within the ceramic and along their interface, calculated through the number of broken bonds per particle. 
For brittle materials, one can then obtain the fracture toughness $K_{IC}$ from the critical energy release rate:
\begin{equation}
    K_{IC}=\sqrt{G_{IC}\dfrac{E_{eff}}{1-\nu^2}},
\end{equation}
where $E_{eff}=(1-f)E_1+fE_2$ is {approximately} the effective Young's modulus of the heterogeneous material for the volume fraction $f$. For further details and discussions on the calculation of fracture toughness for ceramic composites, we refer interested readers to \cite{li2013prediction}.

\begin{figure}[h!]
\centering
\includegraphics[width=.90\columnwidth]{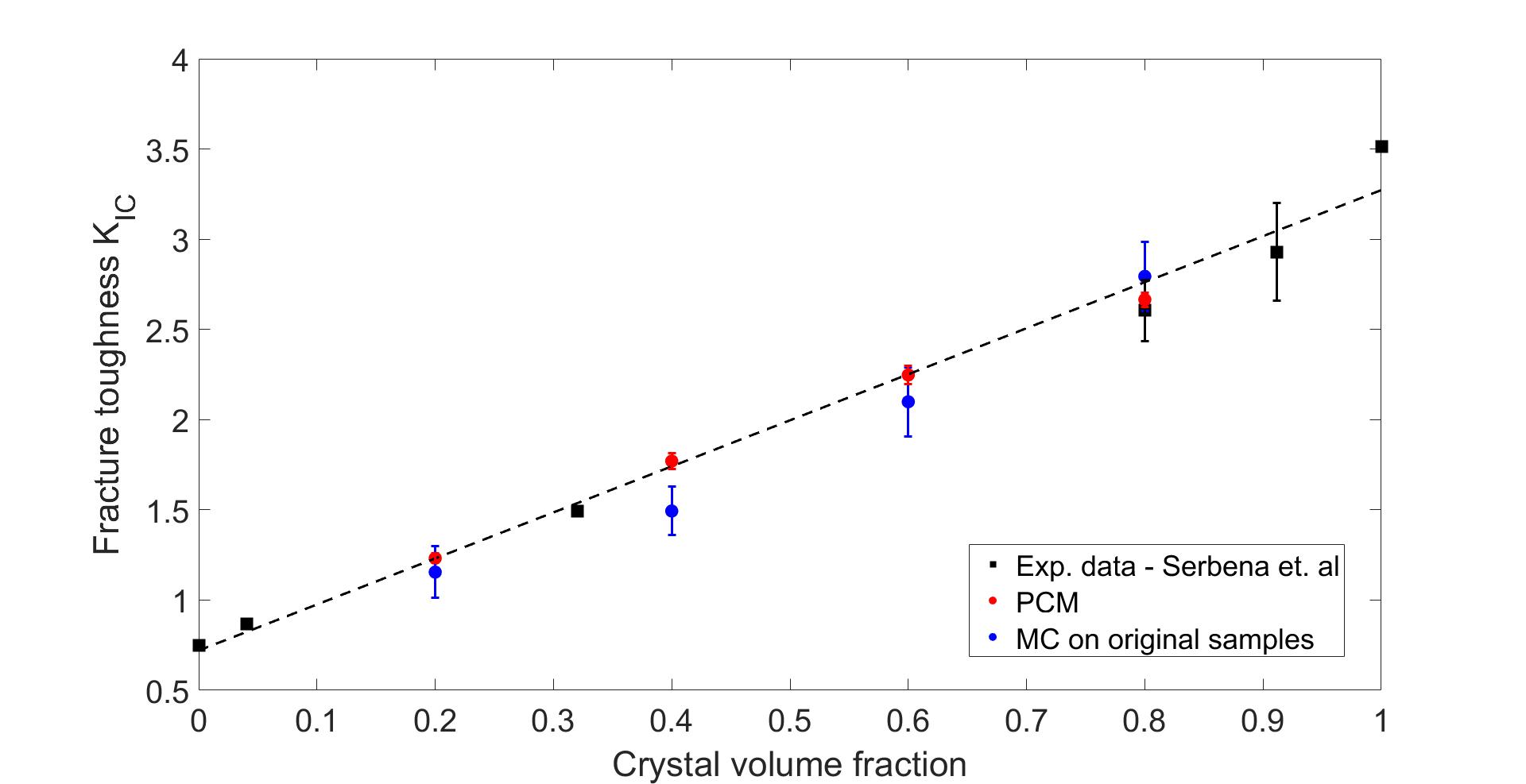}
\caption{Averaged fracture toughness for brittle fracture of glass-ceramics with different crystal volume fractions. Error bars represent standard derivations. Here we use ``PCM'' to denote the results using our proposed probabilistic collocation method approach with $41$ samples from the truncated sampling space, ``MC'' to denote the results using the Monte Carlo method with $100$ samples from the original space $\Omega_p$, and ``Exp'' denotes the experimental results reported in \cite{serbena2015crystallization}.}
\label{fig:KIC}
\end{figure}

Although one can calculate the averaged fracture toughness by sampling $R(\xb,\omega)$ using the Monte Carlo method, we notice that the sampling space $\Omega_p$ is of high dimension and therefore would possibly require a large number of samples. This fact calls for dimensionality reduction for $\Omega_p$ so as to represent the random fields of $E$ and $G$ using a limited number of random variables. In this work the principle component analysis (PCA) approach is employed. In particular, for each crystallized volume fraction value we generate $20,000$ discretized microstrcuture realizations $R(\xb_i,\omega_j)$, $i=1,\cdots,M$, $j=1,\cdots,20000$. Equivalently, we represent each realization by a vector, $\Rb_j\in \real^{M}$, such that $\Rb_j[i]=R(\xb_i,\omega_j)$. 
We then perform PCA to the data matrix formed by $\Rb_j$, $j=1,\cdots,20000$, and keep the first $20$ principle components for dimensionality reduction. To this end, each realization $\Rb_j$ can then be approximated by
\begin{equation}
\Rb_j\approx \overline{\Rb}+\sum_{k=1}^{20} a_{j,k}\Vb_k,\ i=1,2,\dots,M,
\end{equation}
where $\overline{\Rb}$ is the mean of all $\Rb_j$, $\Vb_k$ denote the $k-$th eigenvector in PCA, and $a_{j,k}$ is the $k$-th component of the $j-$th realization. 
Thus, we obtain a truncated representation for the Young's modulus and fracture energy fields in glass-ceramics:
\begin{equation}
E(\xb_i,\xib)= (\overline{\Rb}+\sum_{k=1}^{20} \xi_{(k)}\Vb_k)(E_2-E_1)+E_1,\qquad G(\xb_i,\xib)= (\overline{\Rb}+\sum_{k=1}^{20} \xi_{(k)}\Vb_k)(G_2-G_1)+G_1.
\end{equation}
where $\xi_{(k)}$ is the $k$-th component of $\xib$. We further take $\xi_{(k)}$ as i.i.d. random variables satisfying $\xib_{(k)}\sim \mathcal{N}(0,\lambda_k^2)$, where $\lambda_k$ is the $k_{th}$ eigenvalue in PCA. Noticing that $E$ and $G$ are both affine with respect to $\xib$, and therefore PCM can be applied and the parametric space dimension is $20$. For this example we employ the Smolyak formulation with level $2$, which consists of $41$ samples for each volume fraction value. 
The results are demonstrated in Figure \ref{fig:KIC}, together with the experimental measurements from \cite{serbena2015crystallization}. 
We also report the results using Monte Carlo method as a baseline method, where the fracture toughness for each volume fraction is generated from $100$ realizations from the original sampling space $\Omega_p$. 
From the results, we can observe that the results from both PCA and MC are in good agreement with the experiment data. Comparing between these two methods, although PCA uses less samples, its predictions are more aligned with the linear fitted line from experimental measurements, and are with a low error interval estimation. This validates the applicability of our stochastic LPS solver on providing averaged damage metrics in randomly heterogeneous material fracture problems.

\section{Summary and Discussion}\label{sec:conclusion}
For heterogeneous material modeling problems, different material microstructure, property, interfacial conditions, and operating environments all cause variability within material, which is tremendously difficult to be fully quantified. Therefore, without complete detailed measurements for each individual material sample, it is often non-practical, if not impossible, to provide comprehensive quantitative damage characterization for each sample. This fact calls for stochastic modeling of the variability and characterization of material failure for uncertainty quantification.

In this work, we propose a state-based peridynamics formulation with spatial variability of material properties, to capture the high degrees of complexity and heterogeneity in material damage problems. The well-posedness and convergence to the local problems are studied for the proposed stochastic peridynamics model, which provide a theoretical foundation for numerical developments. An asymptotically compatible meshfree discretization formulation is then developed for the peridynamics model. It provides an efficient representation of interfaces and fracture surfaces. A probabilistic collocation method (PCM) is employed to sample the stochastic process, which guarantees at least algebraic convergence rate for smooth problems in the parametric space, and therefore ensures the sampling efficiency. Therefore, this work has presented a complete workflow demonstrating  how quadrature, heterogeneity and fracture can be handled for linearly elastic materials. In this way, we captures the variability in microstructures and preserves a limit to the relevant local problem as resolution and number of samples are increased. This is a major contribution to the field of peridynamics - while numerous works have demonstrated the flexibility of peridynamics in modeling a diverse set of physical phenomena in a deterministic setting, very few studies have considered the impact of uncertainty in material properties and microstructures. Last but not least, we demonstrate an application of the proposed formulation to estimating the fracture toughness of glass-ceramics, quantitatively validating its applicability in practical engineering problems.

While the current work has been mainly focusing on the physical processes of material damage with uncertainty from material heterogeneity, an important next step is to incorporate other types of uncertainties, such as the variability from interfacial conditions and operating environments. We will additionally consider the generalization of this approach to other types of damage modes, such as the nonlinear elastoplasticity governing ductile failure. As the proposed formulations can be easily extended to 3D problems, we notice that we were unable to perform 3D simulations mainly due to memory limitations of our serial LPS solver. The numerical framework itself is  parallelizable and hence highly scalable, as the meshfree quadrature rule involves only the local construction and inversion of small matrices. In an upcoming work we will investigate how the proposed approach extends to 3D and demonstrate its application in 3D realistic problems.

\section*{Acknowledgements}

Y. Fan, H. You and Y. Yu would like to acknowledge support by the National Science Foundation under award DMS 1753031. Portions of this research were conducted on Lehigh University's Research Computing infrastructure partially supported by NSF Award 2019035. X. Tian's research is supported in part by the National Science Foundation grant DMS-2111608. X. Li's research is supported in part by NSF DMS-1847770 and UNC internal Faculty Research Grants. X. Yang's research is supported in part by the Energy Storage Materials Initiative, which is a Laboratory Directed Research and Development Project at Pacific Northwest National Laboratory. N. Prakash would like to acknowledge discussions with Jason T. Harris, Ross J. Stewart, Binghui Deng, Charlene M. Smith.

\appendix

\section{Truncation Estimates of the Heterogeneous LPS Formulation}\label{app:1}

In this section we provide detailed truncation estimates for the proposed LPS formulations. We first consider the heterogeneous LPS formulation with full Dirichlet-type boundary conditions, proposed in \eqref{eq:nonlocElasticity_comp} and \eqref{eqn:oritheta}. In particular, before showing the proof of Lemma \ref{lem:truncation}, we first show that the nonlocal dilatation $\theta$ is consistent with the local dilatation with the following lemma.

\begin{lemma}\label{thm:thetaconsis} 
Assume that $\ub\in C^4(\overline{\omg\cup\omgbb})$, then there exists $\overline{\delta}>0$ such that for any $0<\delta\leq \overline{\delta}$,   
$$\theta(\xb)-\nabla\cdot\ub(\xb)=D_1\left(\frac{\partial^3 u_1}{\partial x_1^3}(\xb)+\frac{\partial^3 u_2}{\partial x_2^3}(\xb)\right)+3D_2\left(\frac{\partial^3 u_1}{\partial x_1\partial x_2^2}(\xb)+\frac{\partial^3 u_2}{\partial x^2_1\partial x_2}(\xb)\right)+O(\delta^3)=O(\delta^2),$$
for all $\xb\in\omg\cup \omgb$. Here
$$D_1:=\int_{B_\delta (\xb)}K(|\yb-\xb|)(y_1-x_1)^4d\yb= O(\delta^2),\quad D_2:=\int_{B_\delta (\xb)}K(|\yb-\xb|)(y_1-x_1)^2(y_2-x_2)^2d\yb= O(\delta^2).$$
\end{lemma}
\begin{proof}
Denote $\xb=(x_1,x_2)$ where $x_1$ and $x_2$ are the coordinate components along the horizontal and vertical axis, respectively, and $u_1$, $u_2$ as the displacement components along the $x_1$ and $x_2$ directions, respectively. For simplicity, in the following we use $K$ to represent $K(|\yb-\xb|)$ when there is no confusion. For $\ub\in C^4$ and $\xb\in\omg\cup\omgb$, with the symmetry of $B_\delta(\xb)$ we have
\begin{align}
\nonumber&\theta(\xb)-\nabla\cdot\ub(\xb)\\
\nonumber=&O(\delta^3)+\int_{B_\delta (\xb)}K(y_1-x_1)^2\left(\frac{\partial u_1}{\partial x_1}(\xb)+(y_1-x_1)^2\frac{\partial^3 u_1}{\partial x_1^3}(\xb)+3(y_2-x_2)^2\frac{\partial^3 u_1}{\partial x_1\partial x_2^2}(\xb)\right)d\yb\\
\nonumber&+\int_{B_\delta (\xb)}K(y_2-x_2)^2\left(\frac{\partial u_2}{\partial x_2}(\xb)+(y_2-x_2)^2\frac{\partial^3 u_2}{\partial x_2^3}(\xb)+3(y_1-x_1)^2\frac{\partial^3 u_2}{\partial x^2_1\partial x_2}(\xb)\right)d\yb-\dfrac{\partial u_1}{\partial x_1}(\xb)-\dfrac{\partial u_2}{\partial x_2}(\xb)\\
\nonumber=&O(\delta^3)+D_1\left(\frac{\partial^3 u_1}{\partial x_1^3}(\xb)+\frac{\partial^3 u_2}{\partial x_1^2}(\xb)\right)+3D_2\left(\frac{\partial^3 u_1}{\partial x_1\partial x_2^2}(\xb)+\frac{\partial^3 u_2}{\partial x^2_1\partial x_2}(\xb)\right).
\end{align}
\end{proof}

We now proceed to the proof of Lemma \ref{lem:truncation}:
\begin{proof}
We again adopt the coordinate system as in the proof of Lemma \ref{thm:thetaconsis} and denote the two components of $\ub$ as $u_1$ and $u_2$. We notice that 
\begin{equation}\label{eqn:lam_taylor}
\begin{split}
\lambda(\xb,\yb)-\lambda(\xb)=&
\frac{\lambda(\xb)(\lambda(\yb)-\lambda(\xb))}{\lambda(\xb)+\lambda(\yb)}=\frac{\left(\lambda(\yb)-\lambda(\xb)\right)}{2}\left(1+\frac{\left(\lambda(\yb)-\lambda(\xb)\right)}{2\lambda(\xb)}+ O\left(\frac{\left(\lambda(\yb)-\lambda(\xb)\right)}{2\lambda(\xb)}\right)^2\right)\\
=&\frac{\left(\lambda(\yb)-\lambda(\xb)\right)}{2}+O\left( \big(\lambda(\yb)-\lambda(\xb)\big)\right)^2
= \frac{1}{2}\nabla\lambda(\xb)\cdot(\yb-\xb)+O(\delta^2)
\end{split}
\end{equation}
and similarly
\begin{equation}\label{eqn:mu_taylor}
\mu(\xb,\yb)-\mu(\xb)=\frac{1}{2}\nabla\mu(\xb)\cdot(\yb-\xb)+O(\delta^2).
\end{equation}
The bound of $\mcL_{H0}\ub-\mcL_{H\delta} \ub$ can then be obtained via {{Lemma~\ref{thm:thetaconsis}}}, Taylor expansion of $\ub$ and the symmetry of $B_\delta(\xb)$:  
\begin{align*}
\nonumber &\mcL_{H0}(\ub)(\xb)-\mcL_{H\delta} (\ub)(\xb)\\ =&-\dfrac{1}{2}\nabla\cdot[\lambda(\xb) \text{tr}(\nabla \ub(\xb)+(\nabla \ub(\xb))^T)\mathbf{I}+2\mu(\xb)(\nabla \ub(\xb)+(\nabla \ub(\xb))^T)]\\
&+\int_{B_\delta (\xb) }  \left(\lambda(\xb,\yb) - \mu(\xb,\yb)\right)K
     \left(\yb-\xb \right)\left(\nabla\cdot\ub(\xb) + \nabla\cdot\ub(\yb) +D_1\left(\frac{\partial^3 u_1}{\partial x_1^3}(\xb)+\frac{\partial^3 u_2}{\partial x_2^3}(\xb)\right.\right.\\
&\left.\left.+\frac{\partial^3 u_1}{\partial x_1^3}(\yb)+\frac{\partial^3 u_2}{\partial x_2^3}(\yb)\right)+3D_2\left(\frac{\partial^3 u_1}{\partial x_1\partial x_2^2}(\xb)+\frac{\partial^3 u_2}{\partial x^2_1\partial x_2}(\xb)+\frac{\partial^3 u_1}{\partial x_1\partial x_2^2}(\yb)+\frac{\partial^3 u_2}{\partial x^2_1\partial x_2}(\yb)\right)\right) d\yb\\
   \nonumber&+8\int_{B_\delta (\xb) } \mu(\xb,\yb)
     K\frac{\left(\yb-\xb\right)\otimes\left(\yb-\xb\right)}{\left|\yb-\xb\right|^2}  \left(\ub(\yb) - \ub(\xb) \right) d\yb+O(\delta^2)\\
   =&-\dfrac{1}{2}\nabla\cdot[\lambda(\xb) \text{tr}(\nabla \ub(\xb)+(\nabla \ub(\xb))^T)\mathbf{I}+2\mu(\xb)(\nabla \ub(\xb)+(\nabla \ub(\xb))^T)]\\
&+\int_{B_\delta (\xb) }  \left(\lambda(\xb,\yb) - \mu(\xb,\yb)\right)K
     \left(\yb-\xb \right)\left(\nabla\cdot\ub(\xb) + \nabla\cdot\ub(\yb) \right) d\yb\\
   \nonumber&\quad +8\int_{B_\delta (\xb) } \mu(\xb,\yb)
     K\frac{\left(\yb-\xb\right)\otimes\left(\yb-\xb\right)}{\left|\yb-\xb\right|^2}  \left(\ub(\yb) - \ub(\xb) \right) d\yb+O(\delta^2)  
\end{align*}
Hence, by using \eqref{eqn:lam_taylor} and \eqref{eqn:mu_taylor} and their asymptotic orders in terms of $\delta$, and the symmetry of $B_{\delta}(\xb)$, we have
\begin{align*}
&\mcL_{H0}(\ub)(\xb)-\mcL_{H\delta} (\ub)(\xb)\\
=&-\dfrac{1}{2}\lambda(\xb)\nabla\cdot( tr(\nabla \ub(\xb)+(\nabla \ub(\xb))^T)\mathbf{I}-\mu(\xb)\nabla\cdot(\nabla \ub(\xb)+(\nabla \ub(\xb))^T))\\
&\quad \;-(\nabla\cdot \lambda(\xb)\mathbf{I})\nabla\cdot\ub(\xb)
-\nabla\mu(\xb)\cdot (\nabla \ub(\xb)+(\nabla \ub(\xb))^T)\\
      &\; +\left(\lambda(\xb) - \mu(\xb)\right)\int_{B_\delta (\xb) }  K
     \left(\yb-\xb \right)\left(\nabla\cdot\ub(\xb) + \nabla\cdot\ub(\yb)\right) d\yb\\
&\;\;+\frac{1}{2}\int_{B_\delta (\xb) }  \left(\nabla\lambda(\xb)-\nabla\mu(\xb)\right)\cdot(\yb-\xb)K
     \left(\yb-\xb \right)\left(\nabla\cdot\ub(\xb) + \nabla\cdot\ub(\yb)\right) d\yb\\
     &\; +8\mu(\xb)\int_{B_\delta (\xb) } 
     K\frac{\left(\yb-\xb\right)\otimes\left(\yb-\xb\right)}{\left|\yb-\xb\right|^2}  \left(\ub(\yb) - \ub(\xb) \right) d\yb\\
     &\quad +4\int_{B_{\delta}(\xb)}\big(\nabla \mu(\xb)\cdot(\yb-\xb)\big)K \frac{(\yb-\xb)\otimes (\yb-\xb) }{|\yb-\xb|^2}(\ub(\yb)-\ub(\xb)) d\yb \\
=&-(\nabla\cdot \lambda(\xb)\mathbf{I})\nabla\cdot\ub(\xb)-\nabla\mu(\xb)\cdot(\nabla \ub(\xb)+(\nabla \ub(\xb))^T)\\
&\;\;+\left(\nabla\lambda(\xb)-\nabla\mu(\xb)\right)\cdot(\nabla\cdot\ub(\xb))\int_{B_\delta (\xb) }  \left(\begin{array}{c}K(y_1-x_1)^2\\
K(y_2-x_2)^2\end{array}\right)
     d\yb\\
  &+4\int_{B_\delta (\xb) } \big(\nabla\mu(\xb)\cdot(\yb-\xb)\big)
     K\frac{\left(\yb-\xb\right)\otimes\left(\yb-\xb\right)}{\left|\yb-\xb\right|^2}  \left(\ub(\yb) - \ub(\xb) \right) d\yb+O(\delta^2)=O(\delta^2).
\end{align*}
\end{proof}

\bibliographystyle{elsarticle-num}
\bibliography{yyu}

\end{document}